\documentclass[12pt,preprint]{aastex}

\slugcomment{Submitted to the Astrophysical Journal}

\shorttitle{LUMINOSITY AND TEMPERATURE BOOSTS IN MERGING CLUSTERS}
\shortauthors{RANDALL, SARAZIN \& RICKER}

\begin{document}

\title{The Effect of Merger Boosts on the Luminosity, Temperature,
and Inferred Mass Functions of Clusters of Galaxies}

\author{Scott W. Randall, Craig L. Sarazin}

\affil{Department of Astronomy, University of Virginia,
P. O. Box 3818, Charlottesville, VA 22903-0818;
swr3p@virginia.edu,
cls7i@virginia.edu}

\and
\author{Paul M. Ricker}
\affil{Department of Astronomy and Astrophysics, University of
Chicago, 5640 S. Ellis Ave., Chicago, IL 60637;
ricker@flash.uchicago.edu}

\begin{abstract}
In the standard Cold Dark Matter model of structure formation,
massive clusters form via the merger of smaller clusters.
N-body/hydrodynamical simulations of merging galaxy clusters have shown that
mergers can temporarily boost the X-ray luminosity and
temperature of the merged cluster above the equilibrium values for
the merged system.
The cumulative effect of these ``merger
boosts'' will affect the observed X-ray luminosity functions (XLFs)
and temperature functions (TFs) of clusters.
One expects this effect to be most important for the most luminous
and hottest clusters.
XLFs and TFs of clusters provide some of the strongest constraints
on cosmological and large-scale-structure parameters, such as
the mean fluctuation parameter, $\sigma_8$, and the matter density
divided by the critical density, $\Omega_0$.
Merger boosts may bias the values of $\sigma_8$ and $\Omega_0$ inferred from
cluster XLFs and TFs if virial equilibrium is assumed.
We use a semi-analytic technique to estimate the effect of merger
boosts on the X-ray luminosity and temperature functions.
The boosts from individual mergers are derived from N-body/hydrodynamical
simulations of mergers.
The statistics of the merger histories of clusters are determined from
extended Press-Schechter (PS) merger trees.
We find that merger boosts can increase the apparent number of hot,
luminous clusters.
For example, in a Universe with $\Omega_0 = 0.3$
and $\Omega_\Lambda = 0.7$ at a redshift of $z=1$, the number
of clusters with temperatures $T > 10$ keV is increased by a factor of
9.5, and the number of clusters with luminosities $L_X > 5
\times 10^{44} \, h^{-2}$ erg s$^{-1}$ is increased by a factor of 8.9.
We have used our merger-boosted TFs and XLFs to derive the cosmological
structure parameters $\sigma_8$ and $\Omega_0$ by fitting Press-Schechter
equilibrium relations to local ($z=0$) and distant (either $z=0.5$ or
$z=1$) cluster samples.
Merger boosts cause $\sigma_8$ to be overestimated by about 20\%.
The matter density parameter $\Omega_0$ may be underestimated by about 20\%,
although this result is less clear.
If the parameters of the fluctuation spectrum are derived from the
observed TF or XLF (e.g., from a low redshift sample), then this
removes most of the boost effect on $\Omega_0$.
However, larger merger boost effects may appear when cluster XLFs and
TFs are compared to cosmological structure parameters derived by other
techniques (e.g., cosmic microwave background fluctuations or the
brightness of distant supernovae).
\end{abstract}

\keywords{
cosmological parameters ---
galaxies: clusters: general ---
intergalactic medium ---
large-scale structure of the universe ---
X-rays: galaxies: clusters
}

\section{Introduction} \label{sec:intro}

Clusters of galaxies have been widely used to provide
useful constraints on cosmological parameters
(e.g.,
Henry \& Arnaud 1991;
Bahcall \& Fan 1998;
Eke et al.\ 1998;
Borgani et al.\ 2001).
This is in part possible
because there is a well-developed theoretical framework which allows
one to predict the mass function (MF) of clusters of galaxies and its
evolution as a function of the cosmology.
Here, the MF is defined as the number density of clusters as a function of
their mass.
One standard method for predicting the MF is to use
Press-Schechter formalism, originally developed by
Press \& Schechter (1974, hereafter PS), and developed in more detail by
Bond et al.\ (1991)
and
Lacey \& Cole (1993), among others, in combination with the
Cold Dark Matter (CDM) model.
In hierarchical structure formation models like CDM, more massive halos form from the
merger of smaller halos.
Values of cosmological parameters can be estimated by comparing theoretical
models for the MF of clusters with observations.

This technique places the strongest constraints on
$\sigma_8$, the RMS mass fluctuations on a scale of $8 \, h^{-1}$ Mpc
where $h$ is the Hubble constant in units of $100$ km/sec/Mpc, and on
$\Omega_{0} \equiv \bar{\rho} / {\rho_c}$, the ratio of the current
mean matter density to the critical density $\rho_c = 3 H_0^2 / ( 8
\pi G )$ at the current epoch
(e.g., Henry \& Arnaud 1991; Kitayama \& Suto 1996;
Eke et al.\ 1998;
Borgani et al.\ 2001;
Ikebe et al.\ 2002).
Roughly speaking, the present-day abundance of clusters determines
the relationship between $\sigma_8$ and $\Omega_{0}$, while the
evolution of clustering with redshift breaks this degeneracy, although
Reiprich \& B\"ohringer (2001) have been able to constrain
$\Omega_{0}$ using only a local cluster sample by assuming a CDM
spectrum of perturbations whose shape parameter depends on $\Omega_{0}$.
In these determinations, the most massive clusters have the greatest
leverage.
Massive clusters are rare objects, and the abundance of the most
massive clusters is very sensitive to the cosmological parameters.

Unfortunately, the MF of clusters of galaxies cannot be
directly observed (save by gravitational lensing in a relatively small
number of cases).
Generally, it is inferred from the X-ray luminosity function (XLF)
or temperature function (TF), using empirical or semi-empirical relations
between the masses of clusters and their temperatures ($T$) or X-ray
luminosities ($L_X$).
These scaling relations are usually applied under the
assumption that the clusters are dynamically relaxed, although in
reality this cannot always be true since in the CDM model larger
clusters are continually forming via mergers of smaller clusters.

Simulations have shown that if two clusters of comparable mass merge
to form a larger cluster, there is a temporary increase in the
cluster's X-ray luminosity and temperature
(Ricker \& Sarazin 2001;
Ritchie \& Thomas 2002).
If the cluster is observed during this period of boosted luminosity
and temperature, the inferred mass will be larger than the actual mass
of the cluster.
In fact, such objects will be preferentially detected in X-ray flux-limited
samples because they are intrinsically more luminous than equal mass
clusters which are not currently experiencing a merger.
This will be particularly true for X-ray selected, high redshift
clusters.
As a possible examples of this, one of the most
distant X-ray selected clusters observed to date, RXJ1053.7+5735, has a morphology
which may indicate an ongoing merger (Hashimoto et al.\ 2002), as does
the most luminous X-ray selected cluster found to date, RXJ1347.5-1145
(Allen, Schmidt, \& Fabian 2002).

If this X-ray luminosity-temperature (hereafter L-T) boost effect is
sufficiently large, and if mergers occur frequently
enough, then the observed XLF and TF, and hence the inferred MF, will
be different from what they would be if all clusters were dynamically relaxed.
One would expect that merger boosts would affect most strongly the high
luminosity and temperature ends of the XLF and TF.
Since the massive clusters which have very high values of $L_X$ and
$T$ are rare, a small contribution of lower mass clusters with L-T boosts
could strongly affect the statistics.
As we noted above, the abundance of the
hottest and  most luminous clusters (which are usually assumed to be the
most massive clusters)
has a strong influence on the inferred values of the cosmological parameters.
In particular, if merger boosts artificially increase the abundance of
hot, luminous clusters at high redshift, the real value of $\Omega_{0}$
will be underestimated.
Since observations of the numbers of high temperature and luminosity
clusters at moderate and high redshifts have been used to infer that
$\Omega_{0}$ is low
(e.g., Bahcall, Fan, \& Cen 1997;
Borgani et al.\ 2001),
this effect could be important.

In this paper we attempt to quantify the effect of mergers on the
observed XLF and TF, and to determine how the X-ray
L-T boost associated with mergers alters the
values of $\sigma_8$ and $\Omega_{0}$ inferred from observations.
We consider three possible cosmologies:
an ``open'' model ($\Omega_{0} = 0.3$, $\Omega_\Lambda = 0$),
a ``flat'' model ($\Omega_{0} = 0.3, \Omega_{\Lambda} = 0.7$),
and an Einstein-de Sitter (EdS) model ($\Omega_{0} = 1$,
$\Omega_\Lambda = 0$).
Here, $\Omega_\Lambda = \Lambda/3 H_0^2$ and $\Lambda$ is
the cosmological constant.
We use Extended Press-Schechter (EPS) theory
(\S~\ref{sec:basic_ps})
and a Monte Carlo technique to build a collection of merger histories of
clusters (``merger trees,'' \S~\ref{sec:trees}) for a variety of cosmologies.
We then build two
sets of XLFs and TFs from the MF given by the merger trees for each
cosmology by averaging together many merger trees (\S~\ref{sec:XLF+TF}).
One set assumes that clusters are always relaxed when transforming
from mass to temperature or luminosity
(\S\S~\ref{sec:M-T+M-L}, \ref{sec:no_boosts}),
while the other set includes the L-T boost effect
(\S~\ref{sec:with_boosts}).
We ignore such non-gravitational effects as preheating since such
effects are mainly important for lower mass clusters and groups
(Ponman et al.\ 1996),
whereas we choose to focus on massive clusters where the effects of merger
boosts are most important.
We also ignore cooling flows at the centers of clusters, which may be
disrupted by mergers.
The disruption of cooling flows may increase the effect of mergers on
X-ray temperatures, but decrease their effect on X-ray luminosities.
The strength and duration of
the L-T boost is estimated for arbitrary
merging masses and impact parameters by interpolating and extrapolating
from the results of
a series of N-body/hydrodynamical simulations of binary cluster
mergers
(\S~\ref{sec:hydro} and Appendix~B).
The angular momenta or impact parameters for the mergers are drawn from
a distribution based on linear theory
(\S~\ref{sec:impact_param}).
The boosted XLFs and TFs are presented in
\S~\ref{sec:XLF+TF} at a variety of redshifts,
and are compared to unboosted results.
The boosted and unboosted XLFs and TFs at several redshifts are fitted
using a PS mass function and equilibrium relations between the mass and
temperature or X-ray luminosity
(\S~\ref{sec:tree_fit}).
The best-fit parameters, specifically $\sigma_8$ and
$\Omega_{0}$, are compared to the ``actual'' parameters used to
build the trees.
Our conclusions are summarized in \S~\ref{sec:discuss}.

\section{Basic Press-Schechter Theory} \label{sec:basic_ps}

The quantity which is given most directly by theory is the MF of clusters,
$n( M, z)$, which is defined such that $n( M, z ) \, d M$ gives the number of
clusters per unit comoving volume with masses in the range
$M \rightarrow M + dM$.
The MF of
clusters at various redshifts and for various cosmologies can be
derived from numerical simulations of clustering, or from
semi-analytic techniques.  Since we wish to compare the effects of
merger boosts on the properties of an otherwise identical ensemble of
clusters in several cosmologies, it is easier to use the
semi-analytical techniques.  A simple expression for the mass function
of clusters is given by PS formalism.  Comparisons to observations of
clusters and to numerical simulations show that PS provides a good
representation of the statistical properties of clusters, if the PS
parameters are carefully selected (e.g., Lacey \& Cole 1993; Bryan \&
Norman 1998).
Based on large N-body simulations, some authors have suggested other
forms for the mass function of dark matter halos which differ somewhat
from the PS predictions (e.g., Sheth \& Tormen 1999; Jenkins et al.\ 2001).
Since we are mainly interested in determining the relative effects of
merger boosts on the XLF and TF, the exact form of the distribution
function we assume should be unimportant, just so long as the same
model is used both when including the effects of merger boosts and
when ignoring them.
PS formalism assumes that galaxies and clusters grow
by the gravitational instability of initially small amplitude Gaussian
density fluctuations generated by some process in the early Universe.
The fluctuation spectrum is assumed to have larger amplitudes on
smaller scales.  Thus, galaxies and clusters form hierarchically, with
lower mass objects (galaxies and groups of galaxies) forming before
larger clusters.  These smaller objects then merge to form clusters.

According to PS, the mass function of clusters is given by
\begin{equation} \label{eq:PSdensity}
n_{\rm PS} (M,z) \, dM =
\sqrt{ \frac{2}{\pi}} \, \frac{ \overline{\rho}}{M}
\, \frac{\delta_{c}(z)}{\sigma^2 (M) } \,
\left| \frac{d \, \sigma (M) }{d \, M} \right|
\, \exp \left[- \frac{\delta_{c}^{2}(z)}{2\sigma^{2} (M) } \right]
\, dM
\, ,
\end{equation}
where $\overline{\rho}$ is the current mean density of the Universe,
$\sigma(M)$ is the current rms density fluctuation within a sphere of
mean mass $M$,
and $\delta_{c}(z)$ is the critical linear overdensity for a region to
collapse at a redshift $z$.

In CDM models, the initial spectrum of
fluctuations can be calculated for various cosmologies
(Bardeen et al.\ 1986).
Over the range of scales covered by clusters, it is generally sufficient
to consider a power-law spectrum of density perturbations,
which is consistent with these CDM models:
\begin{equation}  \label{eq:sigma}
\sigma (M) = \sigma_{8} \, \left( \frac{M}{M_8} \right) ^{-\alpha} \, ,
\end{equation}
where $\sigma_{8}$ is the present day rms density fluctuation on a scale of
8 $h^{-1}$ Mpc, and
$M_8 = ( 4 \pi / 3 ) ( 8 \, h^{-1} \, {\rm Mpc} )^3 \bar{\rho}$
is the mass contained in a sphere of radius 8 $h^{-1}$ Mpc.
The exponent $\alpha$ is given by $\alpha = (n+3)/6$, where the power
spectrum of fluctuations varies with wavenumber $k$ as $k^n$;
following Bahcall \& Fan (1998), we assume $n=-7/5$, which leads to
$\alpha = 4/15$ throughout.
The normalization of the power spectrum and overall present-day abundance of
clusters is set by $\sigma_8$.
The observed present-day abundance of clusters leads to
$ \sigma_8 \approx 0.6 \Omega_0^{-1/2}$ (e.g., Bahcall \& Fan 1998).
We determine the value of $\sigma_8$ by requiring that our models
reproduce the observed local number density of clusters with
temperature $T = 5$ keV, which we take to be $9
\times 10^{-7} \, h^{3}$ Mpc$^{-3}$ keV$^{-1}$ (Henry \& Arnaud 1991),
although some more recent observations suggest a slightly smaller value
for the local number density of clusters at $T = 5$ keV
(Ikebe et al.\ 2002).
We assume a value of $H_0 = 100 \, h$ km/sec/Mpc, since the results
scale in a simple way with the Hubble constant.
We find $\sigma_8=0.827, 0.834, 0.514$ for the open, flat, and EdS models, respectively.

The evolution of the density of clusters is
encapsulated in the critical over-density $\delta_{c}(z)$ in
equation~(\ref{eq:PSdensity}).
The expressions used for $\delta_{c}(t)$ for the
three cosmologies under consideration are given in Appendix~\ref{ap:delta}.

\section{Merger Trees} \label{sec:trees}

One can determine the merger histories of clusters either moving
forward in time (starting at high redshift and following the merger
history of a set of subclusters) or moving backward in time
(starting at present with a set of clusters, and de-merging them
into subclusters).
We follow the latter approach.
This has several advantages.
First of all, the initial populations of clusters is set
at the present time, where we have the best observational data on
the population of clusters.
Secondly, one can start with rich clusters at present, and we don't need
to follow the evolution of low mass systems.
If the merger histories are computed going forward in time,
one needs to follow the histories of a large ensemble of small systems,
most of which never merge into rich clusters.
Since rich clusters are rare objects, this is quite inefficient.

To generate our merger trees, we follow the EPS method outlined by Lacey
\& Cole (1993).
We begin by writing down the conditional probability that a ``parent''
cluster of mass $M_2$ at a time $t_2$ had a progenitor of mass in the range
$M_1 \rightarrow M_1 + d M_1$ at
some earlier time $t_1$, with $M_2 > M_1$ and $t_2 > t_1$:
\begin{equation} \label{eq:eps_prob}
P(M_1, t_1|M_2, t_2)dM_1 =
\frac{1}{\sqrt{2\pi}}
\frac{M_2}{M_1}
\frac{\delta_{c1}-\delta_{c2}}{(\sigma_{1}^2-\sigma_{2}^2)^{3/2}}
\left| \frac{d\sigma_{1}^2}{dM_1} \right|
\exp
\left[-\frac{(\delta_{c1}-\delta_{c2})^2}{2(\sigma_{1}^2-\sigma_{2}^2)}
\right] dM_1 \, .
\end{equation}
Here $\delta_{c1} \equiv \delta_c (t_1)$ and
$\sigma_{1} \equiv \sigma ( M_1 )$, with
similar definitions for $\delta_{c1}$ and $\sigma_{2}$.
Details of the EPS theory and derivations of
equation~(\ref{eq:eps_prob}) are given elsewhere
(see Lacey \& Cole 1993;
Cohn, Bagla, \& White 2001).

As noted by Lacey \& Cole (1993), the EPS expressions are simplified if
we replace the mass $M$ and time $t$ (or redshift $z$) with the
variables $S \equiv \sigma^2 ( M )$ and $\omega \equiv \delta_c ( t )$.
Note that $S$ decreases as the mass $M$ increases, and that $\omega$
decreases with increasing cosmic time $t$.
Let $K(\Delta S, \Delta \omega) \, d \Delta S$ be the probability that
a cluster had a progenitor with a mass corresponding to
a change in $S$ of $\Delta S = \sigma_{1}^2 - \sigma_{2}^2$
in the range $\Delta S \rightarrow \Delta S + d \Delta S$ at an earlier
time corresponding to $\Delta \omega = \delta_{c1}-\delta_{c2}$.
Then, equation~(\ref{eq:eps_prob}) becomes
\begin{equation} \label{eps_prob2}
K(\Delta S, \Delta \omega) d \Delta S = \frac{1}{\sqrt{2 \pi}}
	\frac{\Delta \omega}{(\Delta S)^{3/2}}
	\exp \left[- \frac{(\Delta \omega)^2}{2 \Delta S} \right] d \Delta S
\, .
\end{equation}
where $\Delta \omega = \delta_{c1}-\delta_{c2}$ and $\Delta S =
\sigma_{1}^2 - \sigma_{2}^2$.

As the mass of the progenitors is decreased ($M_1 \rightarrow 0$), the
total number of progenitors diverges.
On the other hand, the total mass associated with these progenitors
converges (equation~\ref{eq:eps_prob}).
Thus, it is useful to consider only progenitors with masses exceeding
some minimum value $\Delta M_c$.
Mergers involving smaller masses are considered to be part of a
continuous accretion process.
In our merger trees, the masses of clusters increase continuously with
cosmic time as a result of this accretion process (or decrease as we
run the merger trees backwards in time).
However, we do not follow the individual histories of subclusters
with masses $M_1 \le \Delta M_c$.
We chose the value of $\Delta M_c = 10^{12} h^{-1} M_\odot$.

\subsection{Monte Carlo Generation of Merger Trees} \label{sec:make_trees}

We employ a Monte Carlo technique to construct merger trees.
Each tree starts with a cluster with an initial mass $M^0$.
We step each cluster back in time, using a small but finite
time step corresponding to a positive increase $\Delta \omega$.
We adaptively change the step size of $\Delta \omega$ as we run the
trees backward in time by satisfying
the criterion specified by Lacey \& Cole (1993) that
$(\Delta \omega)^2 \la \left| d \ln \sigma^2/d \ln M \right|(\Delta M_c/M) S$,
where $M$ is the mass of the cluster at the current time step and
$\Delta M_c $ is the smallest subcluster we wish to resolve individually.
Specifically, at every step we choose a step size equal to half this
maximum value.  We found that using a smaller timestep equal to 1/10 of this
maximum value did not significantly alter our results.

The cumulative probability distribution of subcluster masses is given by
\begin{equation}  \label{eq:int_prob}
P( <\Delta S, \Delta \omega ) =
\int_{0}^{\Delta S} K (\Delta S^{\prime}, \Delta \omega)
\, d\Delta S^{\prime}
= {\rm erfc} \left( \frac{\Delta
		\omega}{\sqrt{2 \Delta S}} \right) \, ,
\end{equation}
where ${\rm erfc} ( x )$ is the complementary error function.
The cumulative probability distribution is defined such that
$P( <\Delta S, \Delta \omega ) \rightarrow 1$ as
$\Delta S \rightarrow \infty$.
To draw a $\Delta S$ from this probability distribution,
we select a uniformly-distributed random number $r$ between 0 and 1,
and determine the corresponding value of $\Delta S$ by solving numerically
the equation $P( < \Delta S, \Delta \omega ) = r$.
The value of $S_1$ of the progenitor is then $S_1 = S_2 + \Delta S$.
The mass of one of the progenitors is given by $\sigma^2 ( M_1 ) = S_1$,
where $\sigma ( M_1 )$ is given by equation~(\ref{eq:sigma}).
The mass of the other progenitor is
$\Delta M = M_2-M_1$.

\begin{figure}
\plotone{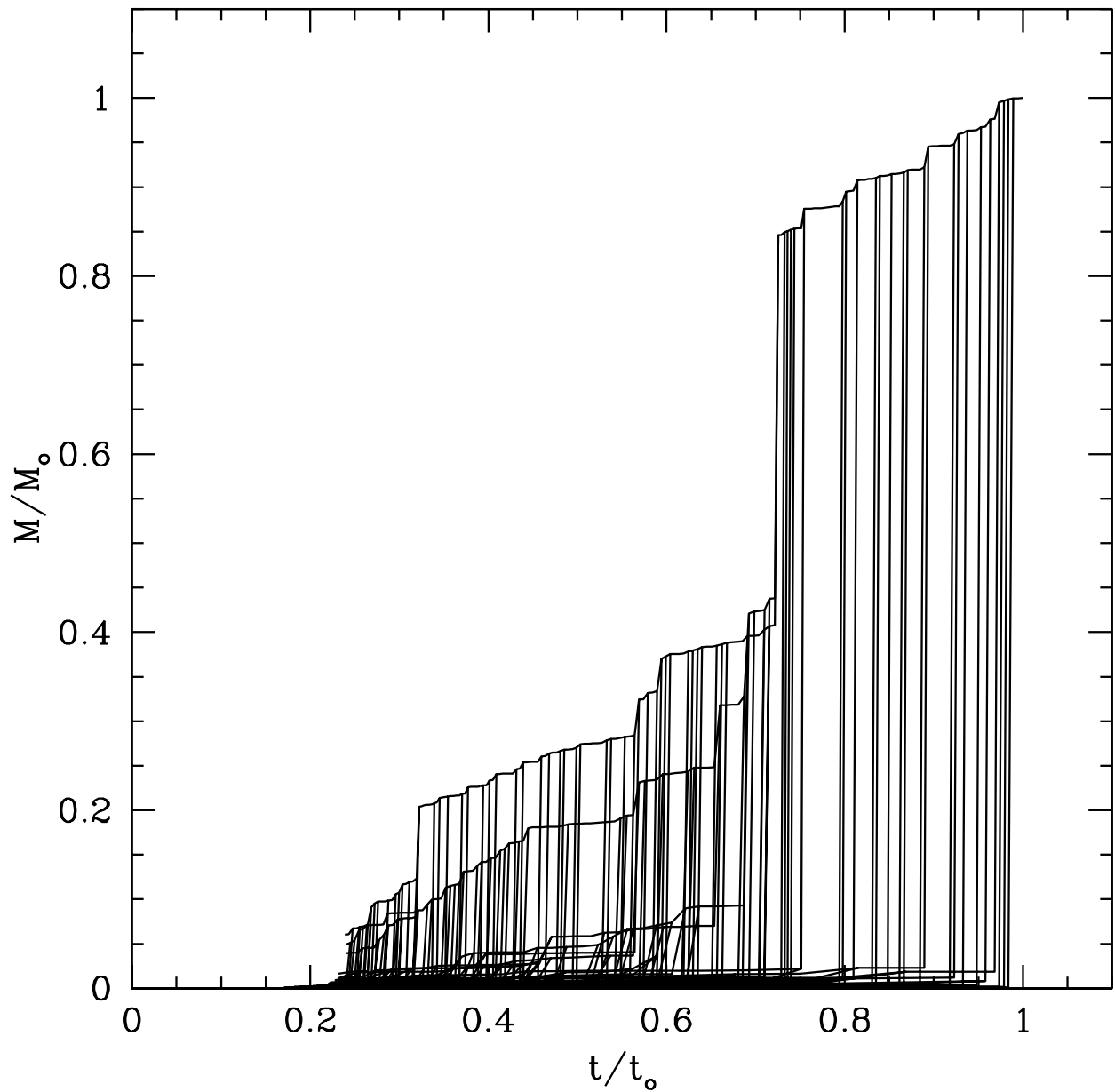}
\caption{Sample merger tree for an open model Universe ($\Omega_0 = 0.3$,
$\Omega_{\Lambda} = 0$).
Time is scaled by the current age of the Universe $t_0$,
and mass is scaled by the mass $M^0$ of the parent
cluster at $z=0$.
For this tree, $M^0 = 10^{15} h^{-1} \, M_\odot$.
\label{fig:merger_tree}}
\end{figure}

Let $M_< \equiv {\rm min}(M_1, \Delta M)$ and $M_> \equiv {\rm max}(M_1, \Delta M)$.
If $M_< \le \Delta M_c$, we consider the change in mass to $M_2$
to be due to accretion (i.e., to a very small merger or mergers).
The transitory effects of such a small merger or mergers are assumed to
be small.
We reduce the mass of the cluster to $M_{\rm new} = M_2 - M_<$, where
$M_{\rm new}$ becomes the new value of $M_2$ for the next timestep in
the merger tree.  We do not follow the smaller mass $M_<$ as a separate subcluster.
However, if $M_< > \Delta M_c $,
we say that a merger has occurred and we follow both of
the merging subclusters of masses $\Delta M$ and $M_1$ separately.
We continue this process until either the branch we are following drops
below a mass of $\Delta M_c$ or we reach the maximum redshift $z_{\rm max}$ we
wish to study.
We choose to run each tree back to a fixed mass limit, as opposed to
a limit expressed as some fraction of the final mass $M^0$, so that
the mass completeness limit of the cluster sample at all redshifts
can be readily determined.
An example of a merger tree generated in this way is shown in Figure \ref{fig:merger_tree}.

As pointed out by Somerville \& Kolatt (1999, hereafter SK), there are some
inherent problems with generating merger trees in this way.
First of all, this technique allows for binary mergers only, while in reality
multiple nearly simultaneous mergers are possible.
SK showed that allowing for multiple mergers improves the agreement between
Monte Carlo merger trees and PS theory predictions for the mass
function.
However, such multiple mergers mainly affect the relatively low mass
halos at rather high redshifts $z \gg 1$.
Since we are mainly interested in massive clusters and will only
follow the merger trees out to $ z \la 2$, this is not such a
significant problem here.

Another related problem with the Binary Merger Tree Method, also pointed out
by SK, is that while we draw one merging halo mass $M_1$ from
equation~(\ref{eq:int_prob}), the mass of the other merging halo
$\Delta M$
is simply taken to be the mass of the progenitor $M_2$ minus $M_1$.
While this approach ensures mass conservation, it distorts the
distribution function we are actually drawing halo masses from in a
complicated way, such that the effective distribution is no longer described
by equation~(\ref{eq:int_prob}).
The solution is to continue to draw merging halo masses from
equation~(\ref{eq:int_prob}) with the restriction that the sum of the
halo masses does not exceed the mass of the progenitor (see SK).
This requires that we allow multiple merger events.

The main reason we have not implemented this N-Branch Tree Method
developed by SK is that it is unclear how to determine the luminosity
and temperature boosts associated with a multiple merger event given
the simulation data available (\S~\ref{sec:hydro}).
However, it is likely that the effects of multiple mergers with smaller
subclusters will be less important than major binary merger events with
$\Delta M \approx M_1$.
In addition,
we find that the predictions of our Binary Merger Tree Method agree
reasonably well with PS theory over the range of redshifts
and halo masses we are interested in (see \S~\ref{sec:tests}).

\subsection{Ensemble of Present-Day Masses} \label{sec:mass_set}

The previous section describes how we determine the merger history
of a cluster with a given mass $M^0$ at the present time.
In order to describe the statistical properties of mergers,
we construct an ensemble of such merger trees for a set of present day
masses, $M^0_i$.
In principle, the simplest way to select such initial masses would
be to draw them from the PS distribution at the present time,
$n_{\rm PS} ( M, z = 0 )$ (equation~\ref{eq:PSdensity}).
Then, the properties of the cluster ensemble at any redshift could be
derived simply by averaging over all of the clusters in the set of
merger trees.
However, the PS mass function diverges at low masses (although the
mass contained in the clusters converges).
This means that most of the merger trees selected in this way would
follow the histories of rather small clusters.
In order to have enough statistics to sample the larger clusters
adequately, one would require a very large number of merger trees.

In order to increase the efficiency and still have adequate statistics
on large clusters, we select the initial (present day) cluster masses
according to a different distribution function which gives a greater
weight to higher mass systems.
We draw the values of $M^0_i$ from the distribution function
$d N / d M^0$, which is defined such that $( d N / d M^0 ) \, d M^0$ is
the number of merger trees with initial masses in the range
$M^0 \rightarrow M^0 + d M^0$.
In practice, we take the initial masses to have logarithmically spaced
masses, which corresponds to
\begin{equation} \label{eq:initial_masses}
\frac{ d \, N }{d \, M^0 } =
\frac{ N_{\rm tree} - 1}{ \ln ( M_u / M_l )}
\,
\frac{ 1 }{ M^0}
\, ,
\end{equation}
where $M_u$ is the largest cluster mass considered, and
$N_{\rm tree}$ is the total number of trees run.
We construct three separate ensembles of merger trees for the three
cosmologies under consideration (the open, flat, and EdS models).  In
each case we choose $N_{\rm tree} = 10^4$, $M_u = 1 \times 10^{16}
\, h^{-1} \, M_\odot$, and a lower limit on the present day mass of clusters
considered $M_l = 1 \times 10^{13} \, h^{-1} \, M_\odot$.

\subsection{Test of the Merger Trees --- the Mass Function}
\label{sec:tests}

Since the method of generating merger trees described above is based
on the EPS theory, we expect that the mass function produced by the merger
trees will agree with the PS function (equation~\ref{eq:PSdensity})
for the same cosmological parameters.
Of course, the tree-generated mass functions will be underestimated
for masses smaller than $M_l$, the smallest initial mass for a merger tree.
The properly-weighted mass function from our ensemble of merger trees would
be given by
\begin{equation} \label{eq:mf_ens1}
n_{\rm tree} ( M , z ) =
\int \left\{ \frac{ n_{\rm PS} ( M^0, z = 0 ) \,
P[ M, t(z) | M^0, t_0 ] \, d M^0 \,
\frac{ d \, N }{d \, M^0 }}{ \frac{ d \, N }{d \, M^0 } } \right\} \, ,
\end{equation}
where $t_0$ is the present age of the Universe, and
the progenitor probability distribution $P[ M, t(z) | M^0, t_0 ]$
is given in equation~(\ref{eq:eps_prob}).
The PS cluster mass function is defined per unit comoving volume, and the
factor of $n_{\rm PS} ( M^0, z = 0 )$ weights each initial cluster and
all of its progenitors by the comoving volume per cluster of mass
$M^0$ in the Universe.

Note that the distribution of initial cluster masses $( d \, N / d \, M^0 )$
cancels between the numerator and the denominator in the integral.
However, we have a discrete set of initial masses,
$M^0_i$ for $i = 1, N_{\rm tree}$.
In the discrete limit, the integral $\int d M^0 ( d \, N / d \, M^0 )$
becomes the sum $\sum_{i=1,N_{\rm tree}}$, and the properly weighted mass
function is
\begin{equation} \label{eq:mf_ens2}
n_{\rm tree} ( M , z ) =
\sum_{i=1,N_{\rm tree}}
\left\{ \frac{ n_{\rm PS} ( M^0_i , z = 0 ) \,
P[ M, t(z) | M^0_i , t_0 ] }{
\left. \frac{ d \, N }{d \, M^0 } \right|_{M^0_i} } \right\} \, .
\end{equation}

In the merger trees, the progenitor probability function
$P[ M, t(z) | M^0_i , t_0 ] $ is determined from a discrete set of
progenitors for each tree.
Let $M^p_{i,j} (z) $ for $j = 1, N^p_i (z) $ be the masses of the $N^p_i$
progenitors (with masses greater than $\Delta M_c$) at redshift $z$ of
the present-day cluster with a mass of $M^0_i$.
We order the progenitor masses such that
$M^p_{i,1} \ge M^p_{i,2} \ge  M^p_{i,3} \dots$
We accumulate the mass function in $N^b$ mass bins whose boundary mass values
are $M^b_k$ for $k = 0, N^b$, with $M^b_0 > M^b_1 > M^b_2 \dots$
Thus, the $k^{th}$ mass bin has a width of
$\Delta M^b_k = M^b_{k-1} - M^b_k$.
Then, the tree-generated, binned mass function at redshift $z$ is given by
\begin{equation} \label{eq:mf_ens3}
n_{\rm tree} ( M^b_k , z ) \, \Delta M^b_k =
\Delta M^b_k \,
\sum_{i=1,N_{\rm tree}}
\left[ \frac{ n_{\rm PS} ( M^0_i , z = 0 ) \, N^p_{i,k} (z) }{
\left. \frac{ d \, N }{d \, M^0 } \right|_{M^0_i} } \right] \, .
\end{equation}
Here, $N^p_{i,k} (z) = j^u_{i,k} (z)  - j^l_{i,k} (z)$ is the number of
progenitors at redshift $z$ of the tree with a present-day mass of $M^0_i$
that lie in the $k^{th}$ mass bin,
$j^u_{i,k} (z)$ is the smallest value of $j$ such that
$M^p_{i,j} (z) \ge M^b_{k-1}$, and
$j^l_{i,k} (z)$ is the largest value of $j$ such that
$M^p_{i,j} (z) < M^b_k$.

\begin{figure}
\plotone{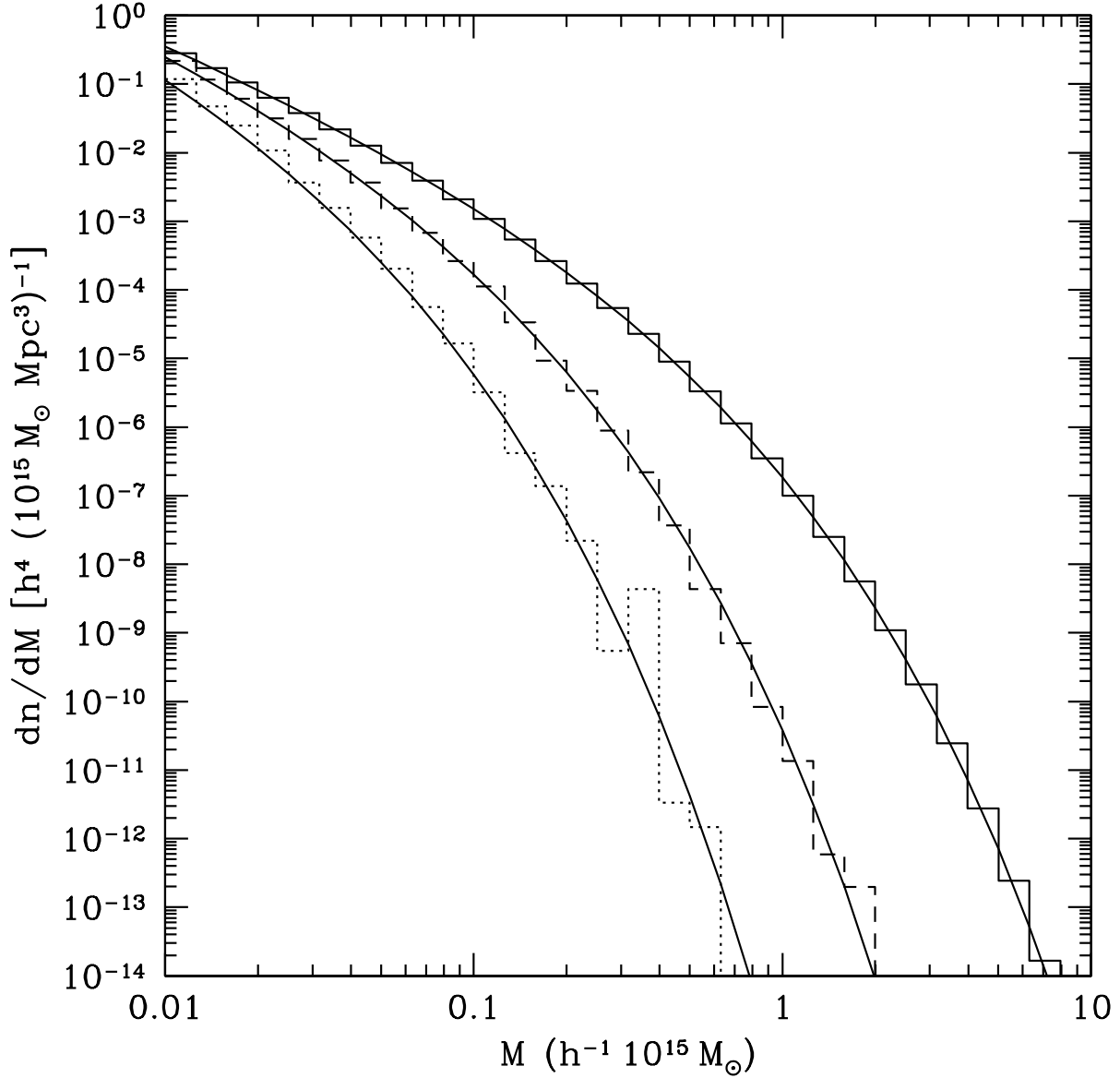}
\caption{Comparison of the mass function given by the merger trees to that
given by the standard PS expression (equation~\protect\ref{eq:PSdensity})
for an EdS Universe.
The histograms show the merger tree mass function for $z=0$ (solid line),
$z=0.5$ (dashed line), and $z=1$ (dotted line).
The smooth curves show the corresponding analytic prediction of the mass
function.
\label{fig:ps_compare}}
\end{figure}

We compared the merger tree prediction of the MF to the PS expression
(equation~\ref{eq:PSdensity}) for the ensemble of trees described in
\S~\ref{sec:mass_set};
the comparison for the EdS model is shown in Figure~\ref{fig:ps_compare}.
If equation~(\ref{eq:PSdensity}) is fit to the mass function generated
by the merger trees allowing $\sigma_8$ and $\Omega_0$ to vary the
values used to construct the trees are recovered to better than 2\% for
$\sigma_8$ and better than 7\% for $\Omega_0$ out to $z=1$.
For the EdS model, the agreement is much better: less than 2\% in
$\sigma_8$ and roughly 2\% in $\Omega_0$.

\section{Non-Boosted Mass-Luminosity and Mass-Temperature Relations}
\label{sec:M-T+M-L}

The merger trees give the masses of clusters as functions of time.  In
order to calculate the TFs and XLFs, we need to convert cluster masses
to temperatures and X-ray luminosities.  We will use the bolometric
X-ray luminosity in order to avoid a specific choice of an energy
band.  In the models including merger L-T boosts, the luminosities
come from hydrodynamical simulations (\S~\ref{sec:hydro}); however,
for comparison to the analytic PS model and to the results of merger
tree simulations without merger L-T boosts, we need a set of
``equilibrium'' or non-boosted relations between the cluster mass and
its X-ray luminosity and temperature.  We could use empirical
relations, but these are generally not available at all redshifts.  We
prefer relations which have the correct dependence on redshift and on
cosmological parameters, since we consider a range of redshifts and a
variety of cosmological models.  Moreover, we prefer a theoretical
dependence on the cluster mass rather than an empirical one, since the
empirical relations may already be biased by the effects of merger
boosts.  It has been shown that the empirical relationship between
cluster mass, luminosity, and temperature is somewhat different from
what one would expect from theoretical arguments and numerical
simulations where only gravitational heating is considered, even at
the relatively high temperatures and luminosities we consider in this
paper (e.g., Arnaud \& Evrard 1999; Allen, Schmidt, \& Fabian 2001;
Finoguenov, Reiprich, \& B\"ohringer 2001).  At present, the detailed
physics and redshift dependence of these non-gravitational effects are
not well understood.  As a result we did not include such effects in
our simulations.  While it is true that preheating and other
non-gravitational effects may slightly dilute the effect of merger
boosts since the fractional increase in X-ray luminosity and
temperature due to mergers will be smaller if non-gravitational
heating is considered, the effect will be most important for low mass
clusters (Ponman et al.\ 1996) which are not of primary interest here.
Since we are mainly interested in massive clusters, are not interested
in these non-gravitational effects, and are primarily concerned with
the relative effects of clusters mergers, it is preferable to use
mass-temperature and mass-luminosity relations which are consistent
with purely gravitational heating.

We adopt the following mass-temperature and mass-luminosity relations
given by Bryan \& Norman (1998):
\begin{equation} \label{eq:M_T}
kT = 1.39 \, N_T \,
\left( \frac{M}{10^{15} \, M_\odot} \right)^{2/3}
\, \left[ h^2 \, \Delta_{c} \, E( z_f )^2 \right]^{1/3} \, {\rm keV} \, ,
\end{equation}
\begin{equation} \label{eq:M_L}
L_{\rm bol} = 1.3 \times 10^{45} \, N_L \,
\left( \frac{M}{10^{15} M_\odot} \right)^{4/3}
\left(  h^2 \Delta_c E( z_f )^2 \right)^{7/6}
\left(  \frac{\Omega_b}{\Omega_o} \right)^2
\, {\rm erg} \, {\rm s}^{-1} \, ,
\end{equation}
where we have added our own normalization factors $N_T$ and $N_L$.
In these equations $\Omega_b$ is the baryon density parameter and $E(z)$ is defined as
$E(z)^2 \equiv \Omega_0 (1+z)^3 + \Omega_{R}(1+z)^2+\Omega_{\Lambda}$, where
$\Omega_{R} \equiv 1/(H_0 R)^2$ and $R$ is the current radius of
curvature of the Universe.
A fit for
$\Delta_c$ is given by Bryan \& Norman (1998) for the relevant
cosmological models:
\begin{equation} \label{eq:Delta_c}
\Delta_{c} = \left\{ \begin{array}{ll}
18\pi^2\,+\,82x\,-\,39x^2  &
(\Omega_{R} = 0) \\
18\pi^2\,+\,60x\,-\,32x^2  & (\Omega_{\Lambda}=0)
			\end{array} \right.
\end{equation}
where $x \equiv [\Omega_0 (1+z)^3/E(z)^2] - 1$.  Since our luminosity
and temperature boosts have been modeled from hydrodynamical simulations
we choose $N_T$ and $N_L$ so that our relations reproduce the
temperature and X-ray luminosity of cluster B in Table 1 of Ricker \&
Sarazin (2001) (assuming cluster B is observed at $z=0$).
In applying these relations to theoretical models of clusters,
the redshift in equations~(\ref{eq:M_T}) and (\ref{eq:M_L}) are taken to
be the formation redshift of the cluster, $z_f$.
Once a cluster is virialized, its properties (including the temperature
and luminosity) are assumed to remain constant within these equilibrium
relations.

\section{Merger L-T Boosts from Numerical Simulations}  \label{sec:hydro}

Using the method described in \S~\ref{sec:make_trees}, we are able to
estimate the number of mergers occurring at a particular redshift.  We
also know the individual masses of the two merging halos.
However, this is all the information we can get from the merger trees
themselves;
if we want to know about the individual cluster X-ray luminosities and
temperatures, and the effect that a merger has on these quantities, we
must turn to other methods.

For non-merging, virialized halos we expect a simple relationship
between mass and temperature or mass and luminosity.
For such halos we assume the relationships given by
equations~(\ref{eq:M_L}) and (\ref{eq:M_T}) throughout.
However, if a cluster has recently experienced a major merger, it will be
in a non-equilibrium state, and we expect its temperature and luminosity to
exceed that of a relaxed cluster of equal mass at the same redshift.
Since we are unaware of any verified analytic prediction of the
L-T boost experienced by two merging clusters of known mass,
we turn to hydrodynamical simulations of two merging clusters.
The general method is to simulate a series of
binary cluster mergers with varying mass ratios and impact parameters
and observe the L-T boost experienced by the merged cluster in each
case, and then to fit a function of the merging masses and the impact
parameter to the observed L-T boosts.  This fitted function is then
used to determine the L-T boost associated with the merger of two
clusters of arbitrary mass at any impact parameter.  The details of
how this fit is carried out and the results are given below in
\S~\ref{sec:hydro_fit}

\subsection{The Simulations}   \label{sec:hydro_info}

The details of the hydrodynamical simulations used to develop a
model for the L-T boost are described in detail elsewhere
(Ricker \& Sarazin 2001).
Data were available from 8 different
simulation runs corresponding to three different mass ratios
$M_>/M_<=1,3,5$
at three different impact parameters
$b= (0,2,5) r_s$,
excluding the one case $M_>/M_< = 5, b = (0,2) r_s$.
Here, $r_s$ is the scale radius of the more massive cluster, which
is the parameter in the assumed Navarro, Frenk, \& White (1997, hereafter
NFW) mass profile of the cluster, and is defined in \S~\ref{sec:hydro_trees}.
The smaller mass $M_<$ was fixed at $2\times10^{14} M_\odot$
for each simulation.
Note that while the runs with $M_>/M_< = 5$ are not
mentioned specifically in Ricker \& Sarazin (2001), the method used to
generate them is identical to that used for the runs described therein.

\subsection{Fitting Simulation Data} \label{sec:hydro_fit}

\begin{figure}
\plottwo{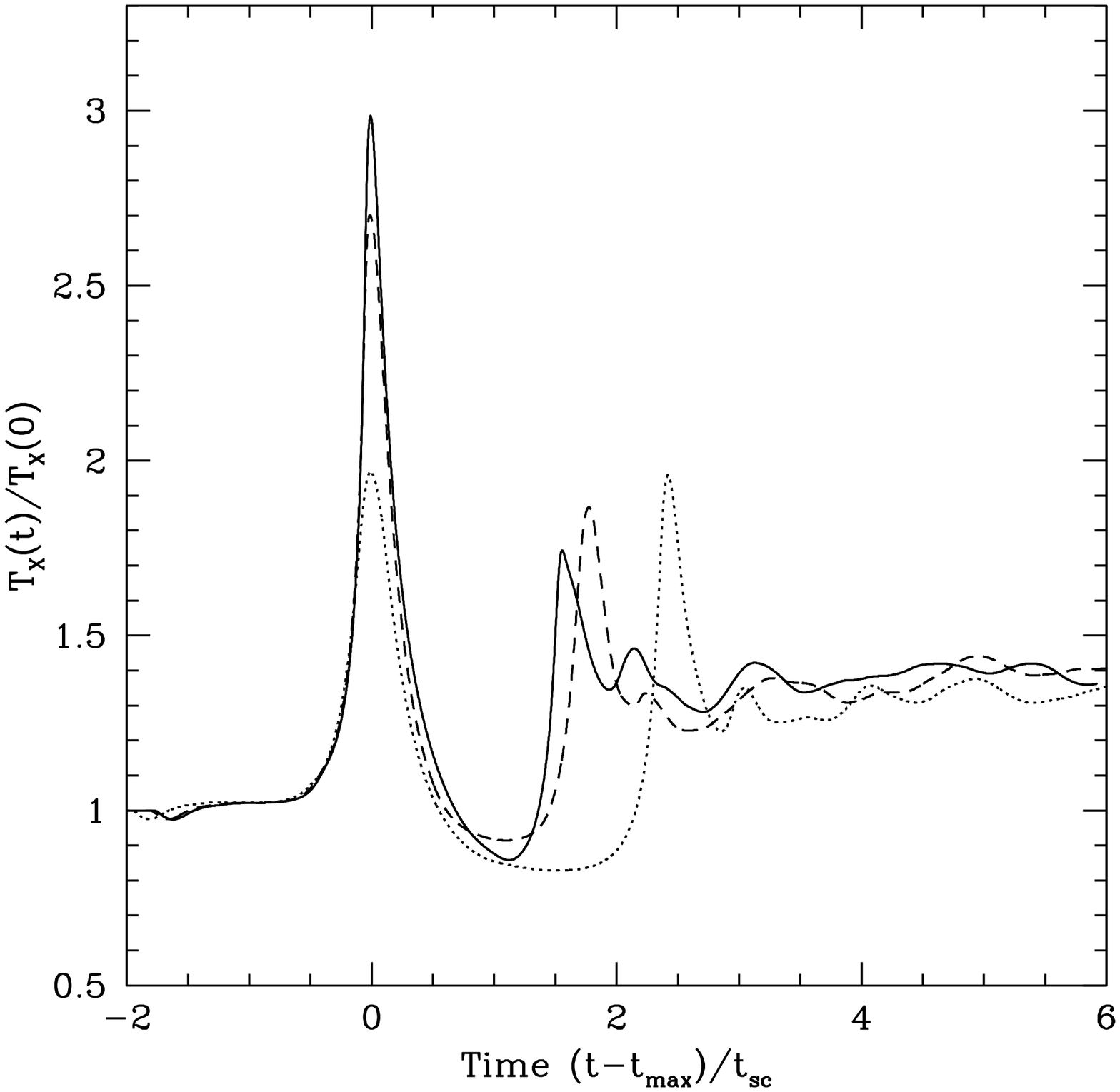}{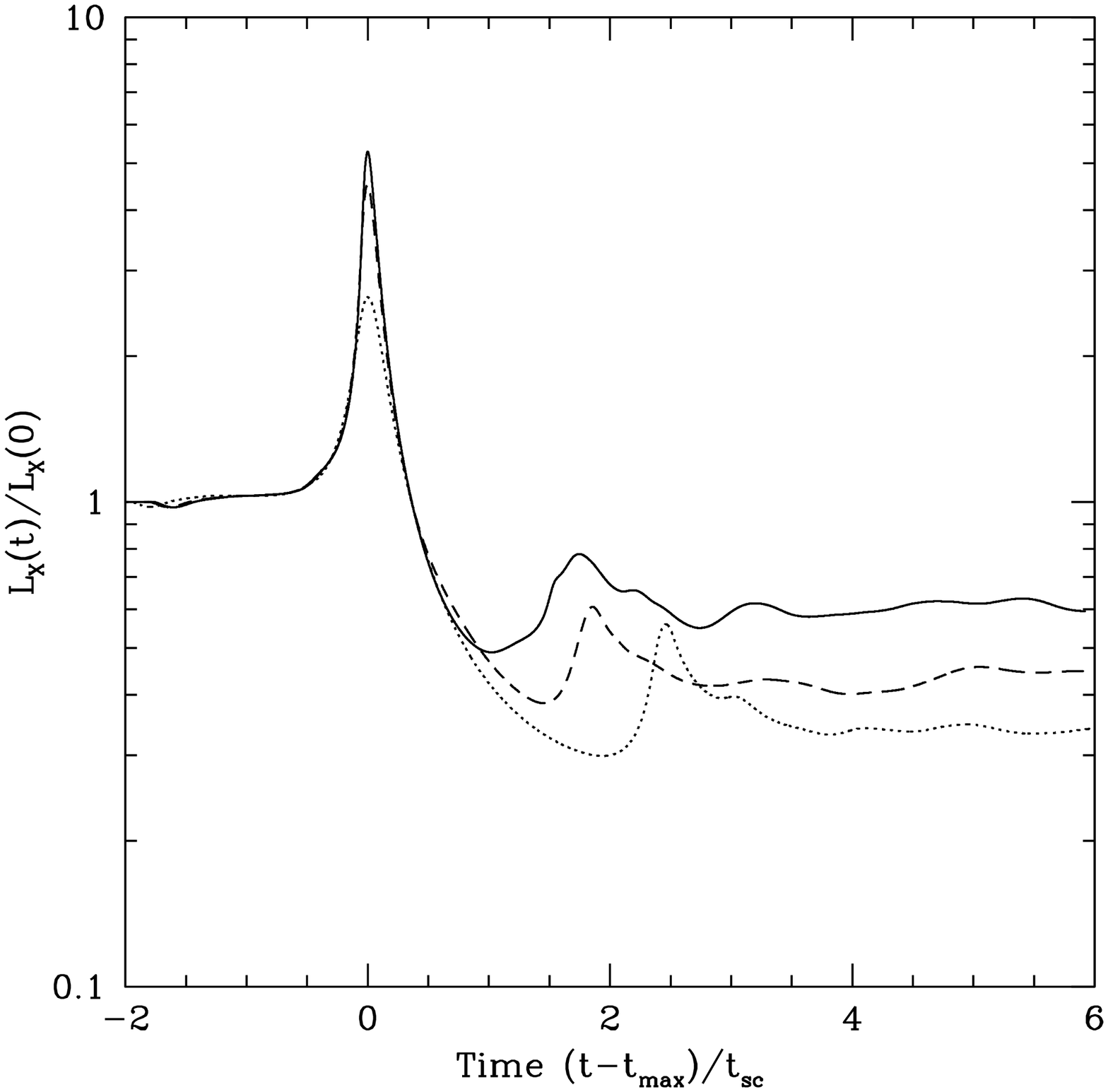}
\caption{Hydrodynamical simulation results on the effects of a merger.
{\it Left:}
Emission-weighted X-ray temperature $T_X$ as a function of time.
{\it Right:}
Bolometric X-ray luminosity $L_X$ (total for both clusters) vs.\ time.
These plots are for equal mass mergers.
The impact parameters for the different runs are $b=0$ (solid line),
$b=2r_s$ (dashed line), and $b=5r_s$ (dotted line), where $r_s$ is the
NFW scale radius of the more massive subcluster.
The time is the offset from the time of peak luminosity
$t_{\rm max}$, scaled by the sound-crossing time $t_{\rm sc}$
of a pre-merger cluster.
\label{fig:LtTt}}
\end{figure}

Figure~\ref{fig:LtTt} shows the behavior of the luminosity and
temperature as a function of time for equal mass merger runs at
different impact parameters
(see also Ricker \& Sarazin [2001], their figures 5 and 8).
$T_X$ is the emission-weighted X-ray temperature of the
merging clusters, while $T_X (0)$ is the same value
for the sum of the two clusters prior to the merger.
Similarly, $L_X$ is the total X-ray luminosity of the merging clusters,
while $L_X (0)$ is the value prior to the merger.
Note that we plot the bolometric X-ray luminosity while
Ricker \& Sarazin (2001) consider the X-ray luminosity in the $2-10$
keV band.
The times are the offset from the time of peak luminosity,
$t_{\rm max}$,
and are scaled by the sound-crossing time $t_{\rm sc}$
of the more massive cluster
($t_{\rm sc}$ is defined in \S~\ref{sec:hydro_trees}).
Several peaks are evident in both the temperature and luminosity
which correspond to the dense cores of the merging clusters interacting
as they pass close by (or through) one another.

In applying these results to merger trees, one complication is that
the merger trees treat the mergers as instantaneous, while the
hydrodynamical simulations follow the mergers over time.
Moreover, the merger trees are computed with a finite time resolutions
(determined indirectly from $\Delta \omega$; see \S~\ref{sec:make_trees}).
Thus, we don't know the precise time or redshift at which each merger
occurred in the merger trees.
On the other hand, we are interested in the statistical effects of mergers;
obviously, numerical hydrodynamical simulations are needed to follow
individual mergers in detail.
Thus, we assume that each merger occurred at a random time within the
time resolution of the merger tree $\Delta \omega$.
In all cases, the time resolution of the merger trees is fine enough that
there are no significant cosmological changes during a time step: for
a merger which forms a 10$^{13} \, h^{-1} \, M_{\odot}$ cluster the
scale factor changes by less than 4\% over one timestep, and for one
which forms a 10$^{14} \, h^{-1} \, M_{\odot}$ cluster it changes
by less than 0.6\%.
We identify the ``time'' of a merger in a merger tree with the time
$t_{\rm max}$ at which the X-ray luminosity boost due to the merger is
maximum (an offset time of zero in Figure~\ref{fig:LtTt}).
Generally, this corresponds to the first time the subcluster cores pass by
(or through) one another.

For the purposes of determining the temperature or luminosity boost
assuming the merger occurs at a random time in the merger tree time
bin, it is useful to consider the distribution of cumulative times as
a function of the boost.
In Figure~\ref{fig:histos}a, we plot the total time $t$ that the
temperature exceeds some boost $t [ > T_X/T_X (0)]$ (scaled by the sound
crossing time $t_{\rm sc}$) vs.\ the relative boost $T_X/T_X (0)$.
The corresponding plots for the luminosity boost are given in
Figure~\ref{fig:histos}b.
Of course, we are most interested in the sections of these plots which
correspond to strong boosts (the right side of the plots).
In Appendix~\ref{ap:hydro_fit}, we fit these cumulative time distributions
for the numerical hydrodynamical simulations of mergers to simple
analytical forms as a function of the masses of the clusters and the
merger impact parameter.
The parameters of these fits are interpolated or extrapolated to
approximate the merger boost for mergers with arbitrary masses and
impact parameters.
We find that, for boosts greater than a factor of 1.5, our fits reproduce the
boosts given by the hydrodynamical simulations with an average accuracy
of better than 2\% for the temperature boosts $T_X / T_X (0)$ and
7\% for the luminosity boosts $L_X / L_X (0)$.
The fits are less accurate for very small boosts, but obviously these have
a very small effect on the XLFs and TFs of clusters.

\subsection{Application to Merger Trees} \label{sec:hydro_trees}

To determine the luminosity and temperature of a cluster at some observation
time $t_{\rm obs}$, we consider all the mergers the cluster has undergone
prior to $t_{\rm obs}$ and the effect each merger will have on the luminosity
and temperature of the cluster at $t_{\rm obs}$.
Recall that at each merger $M_<$ and $M_>$ are the masses of the less massive and more massive
individual merging subclusters respectively (see \S~\ref{sec:make_trees}).
The impact parameter $b$ of the merger is drawn
from the distribution given by linear theory (see
\S~\ref{sec:impact_param} below).
In the fits of the merger boosts from hydrodynamical simulations,
the scaled impact parameter $b^{\prime} = b/(r_{c>} + r_{c<})$ is
used (Appendix~\ref{ap:hydro_fit}),
where $r_{c>}$ and $r_{c<}$ are the core radii of the merging subclusters.
Following Ricker \& Sarazin (2001) and Appendix~\ref{ap:hydro_fit},
we take the core radius of a cluster to be 1/2 of the scaling radius
$r_s$.
We assume a halo concentration parameter $r_{\rm vir} / r_s = 10$
(Ricker \& Sarazin 2001), where the virial radius $r_{\rm vir}$
is given by (Bryan \& Norman 1998):
\begin{equation} \label{eq:r_vir}
r_{\rm vir} = \left( \frac{3M}{4 \, \pi \, \rho_c \, \Delta_c}   \right)^{1/3}.
\end{equation}

The merger boosts depend on the time elapsed since the merger,
$t_{\rm obs} - t_{\rm max}$
(Figure~\ref{fig:LtTt}).
As noted above, we identify the epoch of the merger in the merger
tree with the time $t_{\rm max}$,
and this time is selected at random within the (small)
timestep in which the merger takes place.
The times for the merger boosts are scaled by the
sound crossing time $t_{\rm sc}$ of the more massive subcluster,
\begin{equation} \label{eq:t_sc}
t_{\rm sc} = r_{\rm vir}/c_s \, ,
\end{equation}
where $c_s$ is the sound speed in the cluster,
\begin{equation} \label{eq:c_s}
c_s = \sqrt{\frac{5}{3}\frac{k\,T_X}{\mu \, m_H}} \, .
\end{equation}
The mean mass per particle in the gas is
$\mu = 0.59$ in units of the mass of a hydrogen atom $m_H$.
When computing the sound speed we make the approximation that the
temperature $T_X$ is given by equation~(\ref{eq:M_T}), although note
that Ricker \& Sarazin (2001) use the central temperature when
computing the sound speed.

Equation~(\ref{eq:hyperbola}) is then used with $M_<$, $M_>$,
$b^{\prime}$, and $t_{\rm sc}$
to determine the boost to the X-ray luminosity and temperature
in a cluster at $t_{\rm obs}$ due to a merger at $t_{\rm max}$.
We apply the maximum boost at $t_{\rm obs}$ from all previous
mergers.
Applying these relations
to every cluster in the merger trees at $t_{\rm obs}$ gives a
collection of luminosities and temperatures from which the XLF and the
TF can be constructed as in \S~\ref{sec:no_boosts}.

\section{Angular Momenta and Impact Parameters for Mergers}
\label{sec:impact_param}

The boost in X-ray luminosity and temperature during a merger depends on
the impact parameter or orbital angular momentum for the merger.
Unfortunately, this is not determined by the masses of the
individual subclusters,
and a range of values are possible for mergers of subclusters
with similar masses and similar merger epochs $t_{\rm merge}$.
The merger tree simulations are based on the assumption of spherical
collapse and do not give the impact parameter of the merger.
In hierarchical large scale structure theories, the angular momentum
is generally assumed to arise from tidal torques from surrounding
material.
In linear theory calculations of the growth of perturbations, the
total angular momentum $J$ of a dark matter halo can be characterized
by the spin parameter $\lambda$, defined as
(Peebles 1971)
\begin{equation}
\label{eq:dimensionless spin}
\lambda \equiv {J|E|^{1/2}\over GM^{5/2}}\ .
\end{equation}
Here $E$ is its total energy of the dark matter halo
and $M$ is its mass.
In linear theory, the median value of $\lambda$ is expected to be
approximately constant, independent of the mass of the halo,
and numerical simulations of large scale structure are in reasonable agreement
with this approximation
(e.g.,
Efstathiou \& Jones 1979;
Barnes \& Efstathiou 1987;
Sugerman, Summers, \& Kamionkowski 2000).
The median value of $\lambda$ from linear theory and numerical simulations
is $\lambda_{\rm med} \approx 0.05$.
After these calculations were done we became aware of a more recent
study of the distribution function of a parameter related to
$\lambda$ by Bullock et al.\ (2001).
They propose an empirical log-normal fit to the distribution of $\lambda$
values.
Their simulations and their empirical fit are consistent with the distribution
we have assumed, as long as the median value is fixed at
$\lambda_{\rm med} = 0.05$.
However, their simulations actually give a slightly smaller value of 
$\lambda_{\rm med}$, which would slightly increase the effect of merger
boosts.

Numerical simulations show that individual dark matter halos of the
same mass acquire a range of different values of $\lambda$.
We will assume that the individual components of the angular momenta of
dark matter halos have a Gaussian distribution with a dispersion
$\sigma_\lambda$, which implies that the
total angular momentum $J$ or spin parameter $\lambda$ has a
Maxwell-Boltzmann-like distribution,
\begin{equation}
\label{eq:lambda_dist}
p ( \lambda ) \, d \lambda =
\left( \frac{2}{\pi} \right)^{1/2} \,
\frac{\lambda^2}{\sigma_\lambda^3}
\exp \left( - \frac{\lambda^2}{2 \sigma_\lambda^2} \right) \, d \lambda \, ,
\end{equation}
where $p ( \lambda ) \, d \lambda$ gives the probability that the spin
parameter is  in the range $\lambda \rightarrow \lambda + d \lambda$.
The median value is
$\lambda_{\rm med} = 1.54 \sigma_\lambda$,
which gives $\sigma_\lambda \approx 0.0325$.
Figure~\ref{fig:lambda} shows the distribution function predicted by
equation~(\ref{eq:lambda_dist}) compared to the distribution measured
from simulations by
Bullock et al.\ (2001).
We will assume that this distribution with $\lambda_{\rm med} = 0.5$
applies to the total angular momentum
of the merged clusters, and we draw values of $\lambda$ at random from
this distribution for each merger.

There is a simple kinematic argument which allows one to determine the
impact parameter for a merger from the value of $\lambda$
(Sarazin 2002).
Let $d_0$ be the separation of the subcluster centers at turnaround,
and let $d$ be their present separation.
Then, the impact parameter $b$ is given by
\begin{eqnarray}
b \approx & \lambda \, \sqrt{ \frac{d_0 d}{2}} \,
\left( 1 - \frac{d}{d_0} \right)^{-1/2} \,
f(M_< , M_> )  \nonumber \\
\approx & 160
\left( \frac{\lambda}{0.05} \right)
\left( \frac{d}{1 \, {\rm Mpc}} \right)^{1/2}
\left( \frac{d_0}{5 \, {\rm Mpc}} \right)^{1/2}
\left( 1 - \frac{d}{d_0} \right)^{-1/2}
\left( \frac{f}{2} \right)
{\rm kpc} \, .
\label{eq:b2}
\end{eqnarray}
Here, $ f( M_< ,M_> )$ is a function of the masses of the two subclusters
which is given by
\begin{equation}
\label{eq:fcorr}
f( M_< , M_> ) \equiv \frac{ ( M_> + M_> )^3}
{ M_<^{3/2} M_>^{3/2}} \,
\left[ 1 -
\frac{ \left( M_<^{5/3} + M_>^{5/3} \right)}
{\left( M_< + M_> \right)^{5/3}}
\right]^{3/2} \, ,
\end{equation}
One can show that $ f( M_< , M_> ) \approx 2$ for all masses.

If the two subclusters are treated as point masses and one assumes that
they dominate the mass in their vicinity, then the turnaround separation is
given by
(Sarazin 2002)
\begin{eqnarray}
d_0 & \approx & \left[ 2 G \left( M_< + M_> \right) \right]^{1/3}
\left( \frac{t_{\rm merge}}{\pi} \right)^{2/3} \nonumber \\
& \approx & 4.5
\left( \frac{M_< + M_> }{10^{15} \, M_\odot} \right)^{1/3}
\left( \frac{t_{\rm merge}}{10^{10} \, {\rm yr}} \right)^{2/3}
\, {\rm Mpc} \, .
\label{eqn:init sep}
\end{eqnarray}
Here, $t_{\rm merge}$ is the age of the Universe at the time of the
merger.
In the simulations of Ricker \& Sarazin (2001), the impact parameter is
set at the start of the simulations, when the separation of the two
subclusters is
\begin{equation}
\label{Eqn:initial separation}
d = \sqrt{\left( R_< + R_> \right)^2 + b^2}\ ,
\end{equation}
where $R_<$ and $R_>$ are the cutoff radii of the two clusters
(Ricker \& Sarazin 2001).

\section{Results: Luminosity and Temperature Functions}
\label{sec:XLF+TF}

\subsection{Press-Schechter Non-Boosted Prediction}
\label{sec:XLF+TF_PS}

For the purposes of comparison to our merger tree results, both with and
without merger boosts, it is useful to have a simple prediction of the
XLF and TF based on the PS mass function.
The XLF and TF of virialized, non-merging clusters at an observed redshift
$z$ can be approximated by combining the PS mass function
(equation~\ref{eq:PSdensity}) with the
mass-luminosity (equations~\ref{eq:M_L}) or
mass-temperature (equations~\ref{eq:M_T}) relation.
Note that these relations depend on the redshift $z_f$ at which each
cluster was formed.
Once a cluster is virialized, its temperature and luminosity are assumed
to remain constant.
For clusters of a given mass observed at some redshift $z$, we need to know
the formation redshift $z_f$.
For the merger-tree models, we can calculate $z_f$ from the merger histories.
However, the formation redshift is not uniquely determined by the mass
and observed redshift of a cluster.
A common approximation has been to assume that the formation redshift
is approximately the observed redshift, $z_f \approx z$.
However, this turns out to be inadequate to fit our merger tree results
without merger boosts.

We have taken the formation
redshifts $z_f$ of different clusters of mass $M$ observed at redshift
$z$ to be equal to the mean for each ($M$,$z$), as determined by
Lacey \& Cole (1993) using the parameterization
\begin{equation}
\tilde{\omega_f} \equiv
\frac{\omega_f - \omega}{\left( S_f - S \right)^{1/2}} \, .
\label{eqn:formation redshift}
\end{equation}
The quantities $S$ and $\omega$ were defined in \S~\ref{sec:trees};
$S$ and $\omega$ are the values at the time of the observation, while
$S_f$ and $\omega_f$ are the values at the epoch of the cluster formation.
Cluster formation is assumed to occur when a subcluster has a mass which
is greater than or equal to one half the mass of the cluster at the
time of observation, so that $S_f = S(M/2)$.
So that we will be able to directly compare our analytic
luminosity and temperature functions to the merger tree results,
we adopt a mean value of $\tilde{\omega_f}$ from our own merger trees.
We find a mean value $\langle \tilde{\omega_f} \rangle = 1.18$, which is
in good agreement with the mean of the distribution given by
Lacey \& Cole (1993, see their Figure 9).

Combining equation~(\ref{eq:PSdensity}) with equations~(\ref{eq:M_T})
and~(\ref{eq:M_L}) gives us analytic expressions for the XLF and TF:
\begin{equation} \label{eq:dndT}
n_{\rm PS} (T,z) \, dT =
n_{\rm PS}(M, z) \, \frac{dM}{dT} \, dT =
\frac{3}{2} \, n_{\rm PS}(M, z) \, \frac{M}{T(M, z_f)} \, dT \, ,
\end{equation}
\begin{equation} \label{eq:dndL}
n_{\rm PS} (L,z) \, dL =
n_{\rm PS}(M, z) \, \frac{dM}{dL} \, dL =
\frac{3}{4} \, n_{\rm PS}(M, z) \, \frac{M}{L_{\rm bol}(M, z_f)} \, dL \, .
\end{equation}
The TF is defined so that $n( T, z ) \, d T$ gives the number of
clusters per unit comoving volume with temperatures in the range $T
\rightarrow T + dT$ (similarly for the XLF $n[ L, z ] \, d L$).
The mass-temperature relation $T(M, z_f)$ and mass-luminosity
relation $L_{\rm bol}(M, z_f)$ are given by
equations~(\ref{eq:M_T}) and (\ref{eq:M_L}), respectively.
The formation redshift is determined by solving
equation~(\ref{eqn:formation redshift}) for $\omega_f$ with
$\tilde{\omega_f} = 1.18$, and then converting $\omega_f$ to $z_f$.

\subsection{Merger Tree Prediction Without Merger Boosts}
\label{sec:no_boosts}

The XLF and TF of clusters can be extracted from the merger trees using
the method described in \S~\ref{sec:tests}.
The temperature function in bins $T^b_k$ is given by
equation~(\ref{eq:mf_ens3}) if we replace $M^b_k$ with $T^b_k$, and
if $N^p_{i,k}$ is the number of cluster progenitors with temperatures
in the $k^{th}$ temperature bin.
The luminosity function is given by the same expression, replacing
$M^b_k$ with $L^b_k$, the $k^{th}$ luminosity bin.
We select an observation time or redshift $z$,
identify the collection of progenitor masses which exist in the
merger trees at that time, transform this collection of masses into a
collection of temperatures or luminosities using equation~(\ref{eq:M_T})
or (\ref{eq:M_L}), and then apply the appropriate form of
equation~(\ref{eq:mf_ens3}).
A comparison of the TF given by equation~(\ref{eq:dndT})
to that given by the merger trees is shown in
Figure~\ref{fig:dndT_compare} for an EdS Universe.
The merger tree curves lie slightly above the analytic predictions,
even at $z=0$.
This discrepancy seems to be related to the approximation used to determine
the formation redshifts of clusters when determining the PS TF and XLF
using equations~(\ref{eq:M_T}) and (\ref{eq:M_L});
we assumed $\langle \tilde{\omega_f} \rangle = 1.18$ throughout,
whereas in reality there is a distribution of formation redshifts for
a fixed cluster temperature and luminosity.  In general, this distribution is somewhat
skewed (e.g. Kitayama \& Suto 1996) so that convolving it with the PS
distribution may produce a different TF from the one obtained if an
average formation epoch is assumed.
If, as a test,
we set $z_f = z_{\rm obs}$ in the merger trees and in equation~(\ref{eq:dndT}),
the agreement between the curves in Figure~\ref{fig:dndT_compare} is improved.

\subsection{Merger Tree Predictions With Merger Boosts}
\label{sec:with_boosts}

It is interesting to compare the XLFs and TFs given by the
merger trees when the effects of merger boosts are included to those
given when merger boosts are ignored.  Figures~\ref{fig:dndT_tree}
and~\ref{fig:dndL_tree} plot the boosted and unboosted merger tree
TFs and XLFs, respectively, for the three cosmologies under
consideration.
At low temperatures ($T_X \la 1$ keV) and luminosities
($L_X \la 10^{43} \, h^{-2}$ erg/sec),
there is good agreement between the unboosted and boosted curves.
However, at larger temperatures and luminosities, the boosted XLFs and TFs
rise above the unboosted curves.
This is due to the fact that majors mergers with large boosts are relatively
rare events.
Thus, they do not affect the statistics of clusters unless the virialized
clusters with the same temperature and luminosity are themselves extremely
rare.
There is an exponential drop-off in the expected number of the most massive
clusters (equation~\ref{eq:PSdensity}), which have the largest
virialized temperatures and luminosities.
Thus, relatively rare major mergers of smaller and more common clusters
can compete with the numbers of these most massive clusters.

Table~\ref{tab:boosts} gives the fractional increase in the
\textit{cumulative} luminosity and temperature functions at
$z=0,0.5,1$ due to merger boosts for the three cosmologies we consider.
For example, in the flat model at $z=1$, merger boosts cause
the number of clusters with temperatures $T > 10$ keV to increase by a
factor of 9.49, and the number of clusters with luminosities $L_X > 5
\times 10^{44} \, h^{-2}$ erg s$^{-1}$ is increased by a factor of
8.9.
The same factors for the EdS model at a redshift of $z=0.5$ are 37.7
and 24.1 for the cumulative temperature and luminosity functions respectively.
We note that the increase in the boosted XLFs and TFs over
the unboosted values is largest in the EdS model, weakest in the open
model, and intermediate in the flat model.
This is to be expected since the evolution of cluster abundance is
relatively rapid in the EdS model, less rapid in the flat model, and
least rapid in the open model.
This means that at low redshifts, the merger rate must be highest in the EdS
model and lowest in the open model.
In fact, the evolution of the cluster abundance is so rapid in the EdS model
that the higher temperature and luminosity clusters have essentially
disappeared by a redshift of one.
For this reason, we adapt lower temperature and luminosity limits for
the EdS model at $z=1$ in Table~\ref{tab:boosts}.

It should be noted that at the highest temperatures and luminosities
the merger tree predictions are uncertain for non-zero redshifts since
there are fewer and fewer hot luminous clusters at
larger and larger redshifts, so that the statistics in the high end
bins are poor.  This effect is most noticeable in the EdS model where
cluster abundance evolution occurs most rapidly so that high mass,
high redshift objects are especially rare.  For example, the large
discrepancy at $T \approx 10.5$ keV and $z=1$ between the boosted and unboosted
TFs for the EdS model seen in Figure~\ref{fig:dndT_tree} is not a real
effect but an artifact of the lack of high temperature clusters at
this redshift.

Note that merger boosts cause $T \approx 10$ keV clusters at $z = 1$ in
an EdS model to be almost as common as in the flat model.
Thus, if all of the cosmological parameters other than $\Omega_0$ were
known from other measurements, and the abundance of very hot or very
X-ray luminous clusters at high redshift were used to determine $\Omega_0$,
the value would be substantially underestimated.

\section{Fitting Merger Tree Data} \label{sec:tree_fit}

We now treat the binned XLFs and TFs generated from the merger trees
as we would observational data, fitting the data with
equations~(\ref{eq:dndT}) and (\ref{eq:dndL}) and treating $\Omega_0$
and $\sigma_8$ as free parameters of the fit.  The models are fit
using least squares fitting, where we choose to minimize the square of
the difference of the logs of the PS and merger tree TFs and XLFs so
that more weight is given to hot, luminous clusters.  If we only
consider observations at one redshift, there is a degeneracy between
the fitting parameters $\Omega_0$ and $\sigma_8$, such that $\sigma_8
\approx 0.6 \Omega_0^{-1/2}$ (e.g., Bahcall \& Fan 1998).  One way to
break this degeneracy is to simultaneously fit data at two or more
redshifts.  For each of the three cosmologies under consideration, we
have simultaneously fit the TFs and XLFs at $z=0$ and at either $z =
0.5$ or $z = 1.0$.  This corresponds to the typical procedure of using
observations of a low redshift cluster sample to determine the local
TF or XLF (and possibly infer the mass function), and then comparing
to a higher redshift sample.

The merger boosted TFs and XLFs for the models are shown in
Figures~\ref{fig:dndT_fit} and~\ref{fig:dLdT_fit} along with
best-fit PS models.
The best-fit parameters found in fitting equations~(\ref{eq:dndT})
\& (\ref{eq:dndL}) to the
boosted and unboosted temperature and luminosity functions of the
merger trees are given in
Tables~\ref{tab:temp_fit} \& \ref{tab:lum_fit}.
For comparison, the
actual values of the parameters used to build the merger trees are
also given in these tables.

If we compare the values given for $\sigma_8$ from the boosted TFs and
XLFs to those given by the unboosted functions we see that, in general,
merger boosts cause $\sigma_8$ to be overestimated by about 20\% (except for
the open model XLF, which only shows a 5\% increase).  Similarly, if
we compare the values of $\sigma_8$ from the unboosted TFs to
the values actually used to construct the trees, we find that
they are about 10\% larger.  The same comparison for the XLF results also
show that $\sigma_8$ is larger than the original values used to build
the merger trees, but there is a lot of scatter in the magnitude of
the increase.  The unboosted $\sigma_8$ values are systematically
higher than the original values because the merger
tree TFs and XLFs lie slightly above the curves given by
equations~(\ref{eq:dndT}) \& (\ref{eq:dndL}) with the original values
of $\sigma_8$ and $\Omega_0$.  As mentioned in \S~\ref{sec:no_boosts}, this
has to do with the fact that the formation times
must be estimated when using equation~(\ref{eq:dndT}), whereas for the
merger trees we have a record of the formation time for each cluster.
A similar set of comparisons shows that, in the case of the TFs,
merger boosts cause $\Omega_0$ to be underestimated by about 20\%,
while the unboosted values for $\Omega_0$ are smaller than the actual
values used to build the merger trees by about 15\%.  The XLF results
for $\Omega_0$ show large variations and no systematic trends.  Thus,
while the results from the TFs suggest that merger boosts can cause
$\Omega_0$ to be underestimated by about 20\%, the results from the
XLFs are less conclusive.

It should be noted that the effect of merger boosts on
$\sigma_8$ and $\Omega_0$ depends somewhat on the detailed method used to
determine the TFs and XLFs and the criteria used to fit them.
Thus, the results given here should be viewed as an illustration.
Since the merger boost effect is greatest for the highest temperatures
and luminosities,  the effect on $\sigma_8$ and $\Omega_0$ will be greatest
for samples of hot, luminous clusters, or for fitting techniques that weight
these clusters more highly.

While much of the work done to date on using the XLF and TF of clusters to
constrain both $\Omega_0$ and $\sigma_8$ has used the evolution of these
functions to break the degeneracy between them,
it should be noted that some recent studies simultaneously determine these
two parameters using only the shape of the local mass function of clusters
(e.g., Reiprich \& B\"ohringer 2001).
This study relies on the predicted variation of the fluctuation spectrum
with $\Omega_0$ in CDM models.
Since we assume a power-law fluctuation spectrum, we cannot compare
directly to the results of Reiprich \& B\"ohringer (2001).
However, we did perform a series of fits using only the local $z=0$ data
from the merger trees to determine the relative effects of merger boosts
on the derived cosmological parameters.
The results are consistent with the results obtained by simultaneously fitting
local and higher redshift merger tree data as discussed above.

\section{Discussion and Conclusions} \label{sec:discuss}

Hydrodynamical simulations of binary cluster mergers have shown
that mergers can temporarily boost the X-ray luminosity and
temperature of the merged cluster beyond their virial equilibrium
values.
These merger boosts can alter the observed TFs and XLFs of clusters,
particularly at the high ends.
Even a few ``extra'' clusters with high luminosities and temperatures can have
a significant impact due to the relative rarity of massive clusters.
The mass function of clusters is commonly inferred from the observed TF
or XLF, assuming virial equilibrium.
The inferred mass function is often used to constrain cosmological parameters.
Thus, merger boosts can affect the inferred cosmological parameters.

We tested the amplitude of the merger boost effect by using EPS theory and
a Monte Carlo technique to numerically reconstruct the merger histories
(trees) of a population of clusters.
At each merger, we quantified the strength of the merger boost by
extrapolating or interpolating from a set of hydrodynamical simulations
of cluster mergers.
We then built TFs and XLFs from the merger trees,
averaging together our sets of merger histories
so that the sample cluster population was representative of the
observed present day population.

Our results show that merger boosts do indeed affect the apparent
number of high mass clusters.
For example, in the flat model at $z=1$ merger boosts cause
the number of clusters with temperatures $T > 10$ keV to increase by a
factor of 9.5, and the number of clusters with luminosities $L_X > 5
\times 10^{44} \, h^{-2}$ erg s$^{-1}$ is increased by a factor of
8.9.  The effect is strongest for an EdS model, since clusters evolve
more rapidly in this model than in the open and flat models.  In the
EdS model at a
redshift of $z=0.5$, the number of clusters with temperatures $T > 10$
keV is increased by a factor of 38, and the number with luminosities
$L_X > 5 \times 10^{44} \, h^{-2}$ erg s$^{-1}$ by a factor of 24.
At first this might appear contradictory with the fact that the X-ray
luminosity receives a larger boost from mergers than the X-ray
temperature (see Figure~\ref{fig:LtTt}).  However, the range of X-ray
luminosities is larger than the range of X-ray temperatures; another
way of saying this is that the X-ray luminosity-temperature relationship
is much steeper than linear.
At a result, the overall effect of luminosity boosts on the XLF is not
as strong as the effect of temperature boosts on the TF.

Comparing the boosted and unboosted differential TF for the
EdS and flat models at $z=1$ shows that
merger boosts cause $T \approx 10$ keV clusters to be almost as common
in the EdS model as in the flat model.  This means that if all
cosmological parameters other than $\Omega_0$ were known from other
measurements, and the abundance of very hot or very
X-ray luminous clusters at high redshift were used to determine $\Omega_0$,
the value would be substantially underestimated.

We fit PS distributions to our simulated TFs and XLFs.
We did this by considering two samples:
a low redshift, local sample ($z = 0$),
and a moderate or high redshift sample ($z = 0.5$ or $z = 1$).
The two samples are fit simultaneously to determine
the best-fit values of $\sigma_8$ and $\Omega_0$, and
these values were compared to the actual values used to
construct the merger trees.
Merger boosts can cause $\sigma_8$
to be overestimated by about 20\%.
Merger boosts may also cause
$\Omega_0$ to be underestimated by about 20\%, although the results
from the XLF fits do not show a clear trend.
The effect of merger boosts on the inferred value of $\Omega_0$ is
much smaller in these fits than might have been expected from the
overall increase in the number of hot clusters at high redshifts produced
by merger boosts.
Much of the effect of merger boost is ``renormalized'' away by the
joint fit of low and high redshift sample.
One way to think of this is that merger boosts increase the numbers of
hot clusters, both at low redshift and high redshift.
Determining $\sigma_8$ from the low redshift sample or from a joint fit
removes much of the boost effect, and the change in $\Omega_0$ is smaller
than one might have expected.


Studies of the abundance of hot or luminous clusters at high redshift
have been used to argue that we live in a low density Universe
(e.g., Bahcall \& Fan 1998;
Borgani et al.\ 2001).
It would appear that merger boosts do not invalidate this conclusion,
although the error bars should be increased significantly to include
the systematic uncertainties associated with merger boosts.

On the other hand, merger boosts do affect the value of $\sigma_8$.
As noted before, the number of the hottest and most luminous clusters
are affected quite strongly.
Thus, any comparison between cluster TFs and XLFs and cosmological
parameters derived from other objects (from cosmic microwave background
radiation fluctuations, or from the brightness of distant supernovae)
is likely to be inconsistent and may lead to errors in the deduced
cosmological parameters.
At the least, the systematic uncertainties are likely to be much larger
than might be inferred from the statistics of clusters alone.

Obviously, the effect of merger boosts could be avoided entirely if the
mass function could be determined directly from gravitational lensing
observations.
Radio detections of the Sunyaev-Zel'dovich (SZ) effect can also provide
a measure of the number of luminous and hot clusters, particularly at
high redshift
(e.g., Holder et al.\ 2000).
Although mergers should also boost the microwave decrement from clusters,
we expect that this effect would be smaller than the effect on the
X-ray emission-weighted temperature or the X-ray luminosity, because
the SZ effect depends on density rather than the square of the density.

\acknowledgments

We thank Y. Fujita, T. Reiprich, and M. Takizawa for useful discussions.
S. W. R.  and C. L. S. were supported
in part by $Chandra$ Award Number GO1-2123X, issued by the $Chandra$
X-ray Observatory Center, which is operated by the Smithsonian
Astrophysical Observatory for and on behalf of NASA under contract
NAS8-39073, and by NASA XMM Grant NAG5-10075.  P. M. R. is supported by
DOE under Grant B341495 to the ASCI Flash Center at the University of
Chicago.

\appendix
\section{Cosmological Dependence of $\delta_c$} \label{ap:delta}

The critical overdensity as a function of cosmic time $t$ depends on
the cosmological parameters $\Omega_m$ and $\Omega_{\Lambda}$.  In
this paper we use the following expressions for $\delta_c$:
\begin{equation}   \label{eq:delta}
\delta_{c}(z) = \left\{
\begin{array}{ll}
\frac{3}{2} D( t_0 ) \left[ 1 +
\left( \frac{t_{\Omega}}{t}
\right)^{\frac{2}{3}}
\right]   &
(\Omega_0 < 1 \, , \, \Omega_\Lambda = 0) \\
\frac{3}{2} D( t_0 ) \left[
\left( \frac{t^{\prime}_{\Omega}}{t}
\right)^{\frac{2}{3}} - 1
\right]   &
(\Omega_0 > 1 \, , \, \Omega_\Lambda = 0) \\
\frac{3(12\pi)^\frac{2}{3}}{20} \left(
\frac{t_0 }{t}
\right)^{\frac{2}{3}} &
(\Omega_0  = 1 \, , \, \Omega_\Lambda = 0) \\
\frac{D(t_0 )}{D(t)}
\left(\frac{3(12\pi)^\frac{2}{3}}{20}\right)
\left(1 + 0.0123 \ \log_{10}\Omega_{f}
\right) &
(\Omega_0 + \Omega_{\Lambda} = 1)
\end{array}
\right.
\end{equation}
For the open model ($\Omega_0 < 1$, $\Omega_\Lambda = 0$),
$t_{\Omega} \equiv
\pi H_0^{-1} \Omega_0 \left( 1 - \Omega_0 \right)^{-\frac{3}{2}}$
represents the epoch at which a nearly constant expansion takes over
and no new clustering can occur, and the growth factor can be expressed as
\begin{equation}  \label{eq:growth_factor}
D(t)=\frac{3\ \sinh \eta \left( \sinh \eta - \eta \right)}
	  {\left( \cosh \eta - 1 \right)^{2}} - 2
\end{equation}
where $\eta$ is the standard parameter in the cosmic expansion equations
(Peebles 1980, eqn.~13.10)
\begin{equation} \label{eq:eta}
\begin{array}{ll}
\frac{1}{1+z} = \frac{\Omega_0}
{2 \left( 1 - \Omega_0 \right)}
\left( \cosh \eta - 1 \right) \, , \\
H_0 t = \frac{\Omega_0 }
{2 \left( 1 - \Omega_0 \right)^{\frac{3}{2}}}
\left( \sinh \eta - \eta \right) \, .
\end{array}
\end{equation}
A similar set of relations holds for the closed model ($\Omega_0 > 1$,
$\Omega_\Lambda = 0$):
\begin{equation} \label{eq:closed_growth}
D(t)= 2 - \frac{3\ \sin \eta^\prime \left( \eta^\prime - \sin \eta^\prime \right)}
	  {\left( 1 - \cos \eta^\prime  \right)^{2}},
\end{equation}
\begin{equation} \label{eq:closed_eta}
\begin{array}{ll}
\frac{1}{1+z} = \frac{\Omega_0}
{2 \left( \Omega_0 - 1 \right)}
\left( 1 - \cos \eta^\prime \right) \, , \\
H_0 t = \frac{\Omega_0 }
{2 \left( \Omega_0 - 1 \right)^{\frac{3}{2}}}
\left( \eta^\prime - \sin \eta^\prime \right) \, .
\end{array}
\end{equation}
In this model $t^{\prime}_{\Omega} \equiv \pi H_0^{-1} \Omega_0 \left(
\Omega_0 - 1 \right)^{-\frac{3}{2}}$ represents the lifetime of the
Universe, that is to say the cosmic time at which the Universe
formally recollapses to a singularity.
Although we do not consider a closed cosmological model when building
merger trees it is necessary that our fitting routines be able to
consider cosmologies with $\Omega_0 > 1$ when fitting cosmological
parameters to the merger tree data (see \S~\ref{sec:tree_fit}).
The solution for $\delta_{c}$ in the Einstein-de Sitter model
can be obtained from the open model solution by the limit
$t_{\Omega}/t \rightarrow \infty$
(Lacey \& Cole 1993).
To evaluate $\delta_{c}$ in the flat model
$(\Omega_0 + \Omega_{\Lambda} = 1$),
we have used an approximation given by Kitayama \& Suto (1996).
Here $\Omega_{f}$ is the value of the mass density ratio $\Omega$ at
the redshift of formation,
\begin{equation} \label{eq:omega_f}
\Omega_{f} = \frac{\Omega_0
	\left( 1+z \right)^{3}}
	{\Omega_0 \left( 1+z \right)^{3} + \Omega_{\Lambda}} \, .
\end{equation}
In this model the growth factor can be written as
\begin{equation} \label{eq:growth_flat}
D(x)= \frac{(x^{3} + 2)^{1/2}}{x^{3/2}} \,
\int_{0}^{x} x^{3/2} \, (x^{3}+2)^{-3/2}dx
\end{equation}
(Peebles 1980, eqn.~13.6) where $x_0 \equiv ( \frac{2
\Omega_{\Lambda}}{\Omega_0} )^{1/3} $ and $x = x_0/(1+z)$.

\section{Fitting Simulation Data} \label{ap:hydro_fit}

\begin{figure}
\plottwo{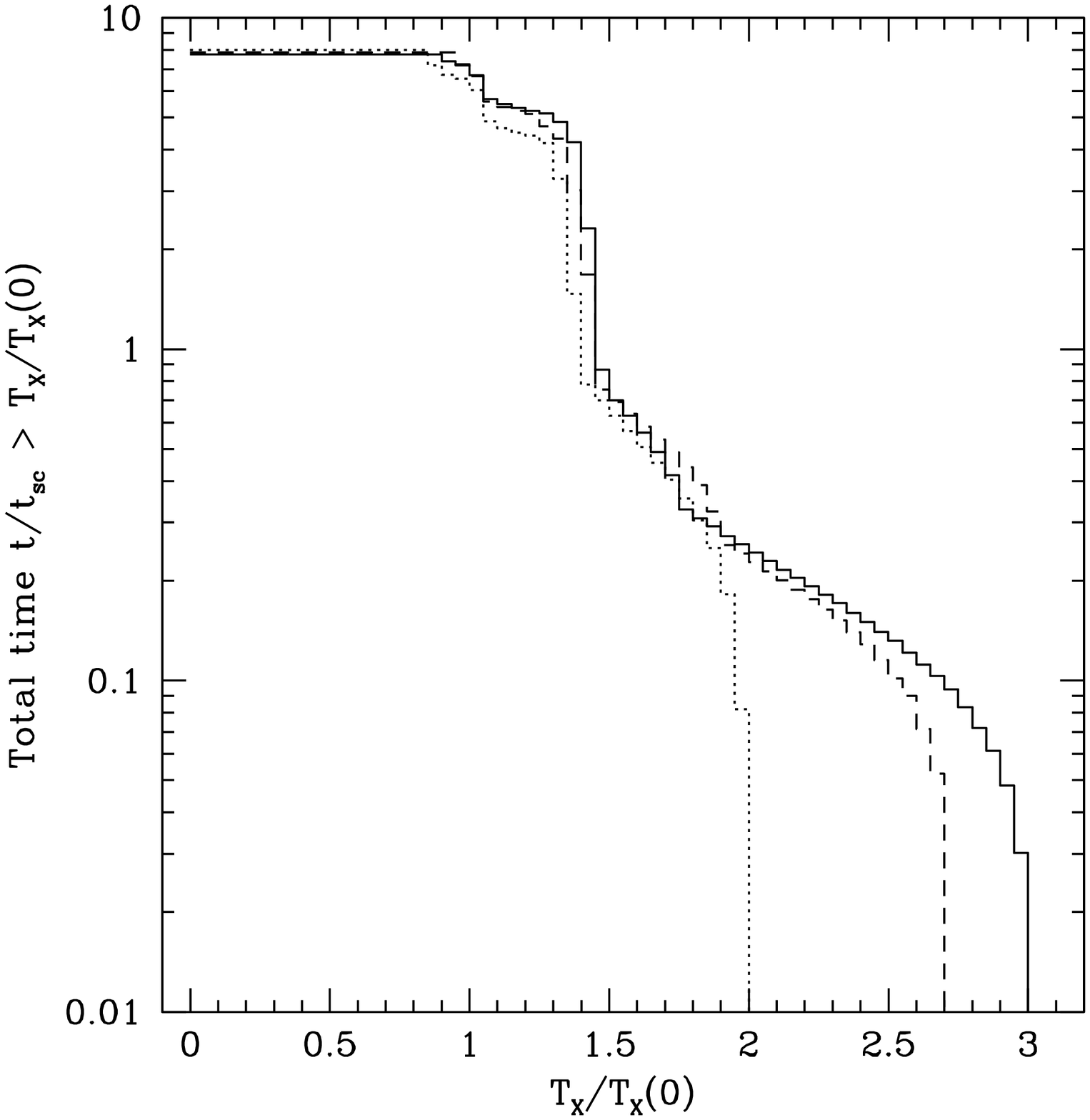}{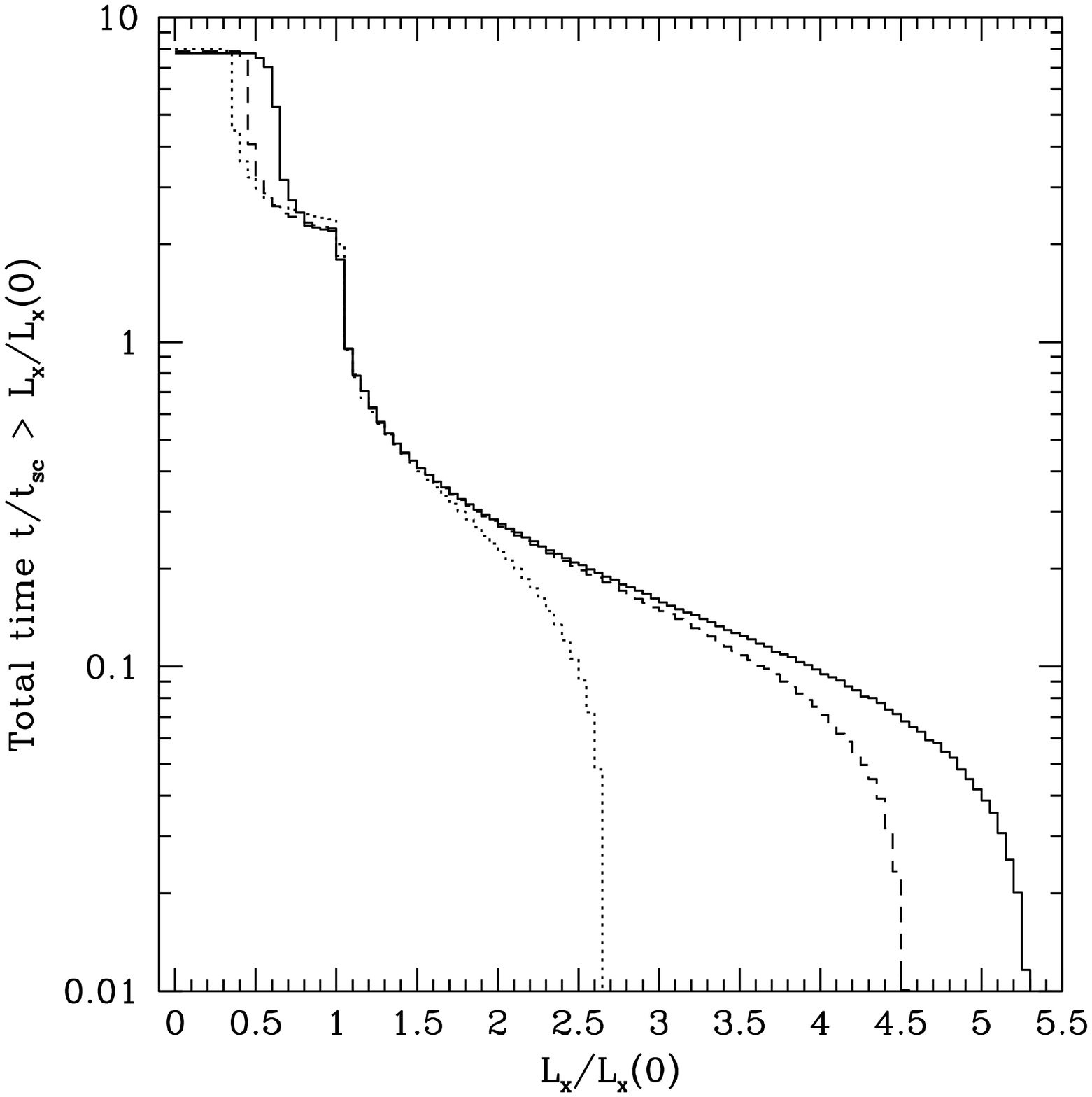}
\caption{{\it Left:}\ The total time $t$ that the emission-weighted
temperature $T_X$ is
boosted above some fraction of its initial pre-merger value $T_X(0)$,
scaled by the sound-crossing time $t_{\rm sc}$ of the more massive cluster.
These histograms were
generated from the hydrodynamical results (Figure~\protect\ref{fig:LtTt}).
Plots are given for the equal mass merger runs, with each cluster mass being
$M = 1.99 \times 10^{14} \, M_\odot$,
for collisions with impact parameters
$b=0$ (solid line),
$b=2\,r_{s}$ (dashed line),
and $b=5\,r_s$ (dotted line),
where $r_s$ is the NFW scaling radius of the more massive cluster.
{\it Right:}\ Same as above, but for the bolometric X-ray luminosity boost
instead of the temperature boost.
\label{fig:histos}}
\end{figure}

\begin{figure}
\vskip4.0truein
\includegraphics{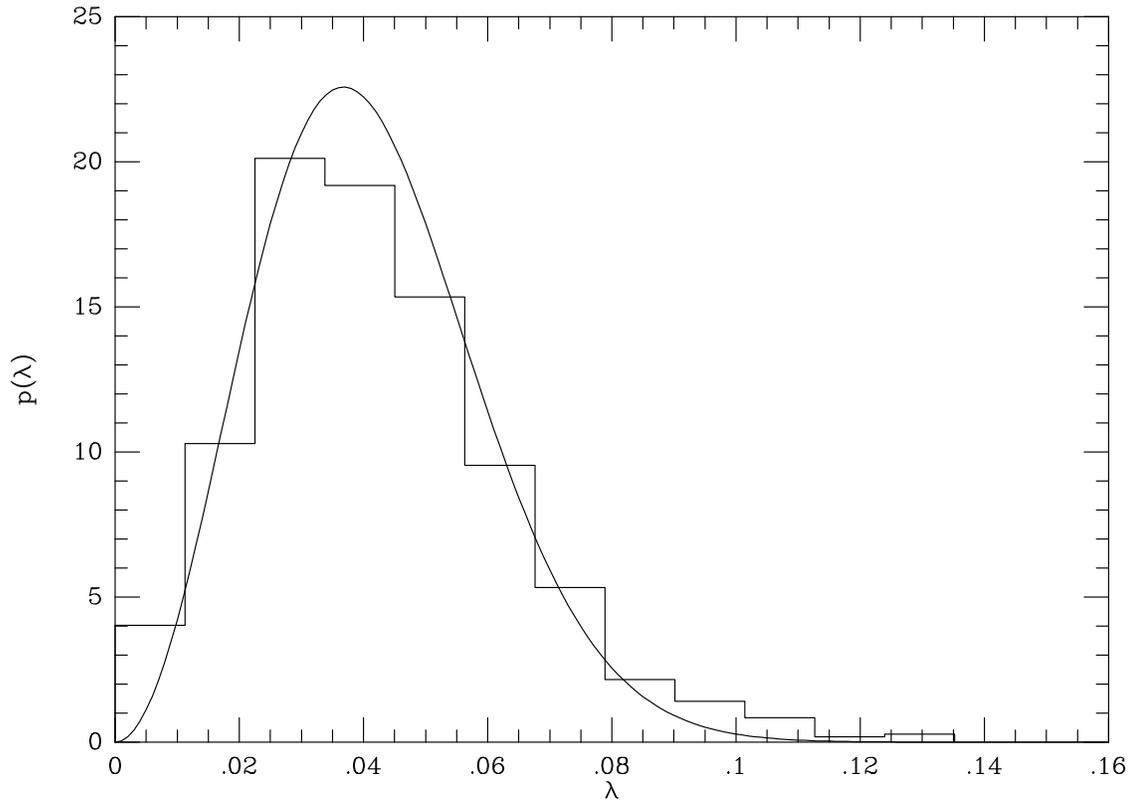}
\caption{The distribution function for the dimensionless spin parameter
$\lambda$.
The smooth curve shows the distribution given by
equation~(\protect\ref{eq:lambda_dist}), while the histogram shows results
from simulations by
Bullock et al.\ (2001).
\label{fig:lambda}}
\end{figure}

\begin{figure}
\plotone{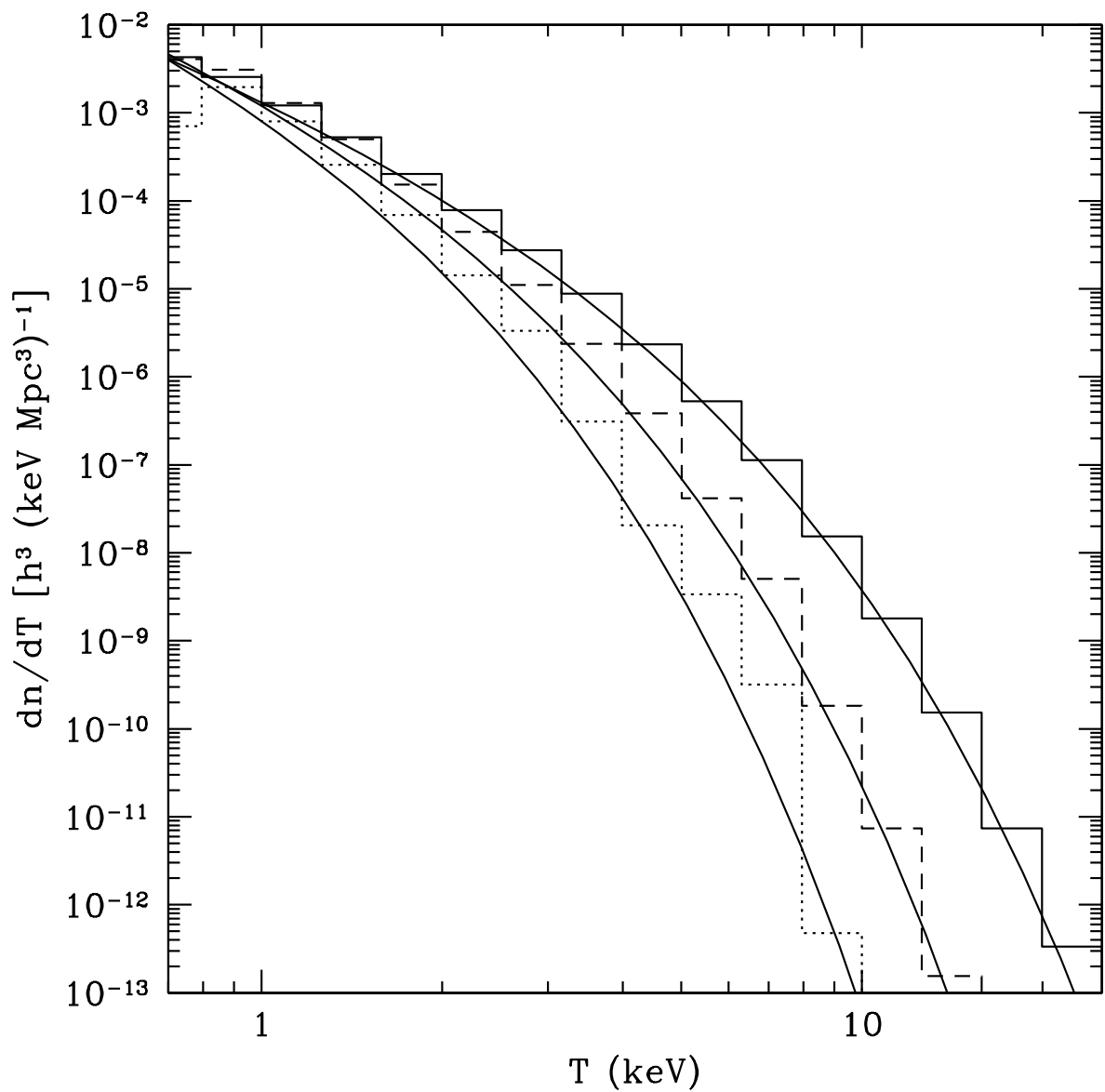}
\caption{Comparison of the non-boosted equilibrium temperature function given
by the merger trees to the PS prediction (equation~\protect\ref{eq:dndT}) for
an EdS Universe.
The histograms show the merger tree TF for $z=0$ (solid line),
$z=0.5$ (dashed line), and $z=1$ (dotted line).   The smooth curves show the
corresponding analytic prediction of the mass function.
\label{fig:dndT_compare}}
\end{figure}

\begin{figure}
\vskip3truein
\includegraphics{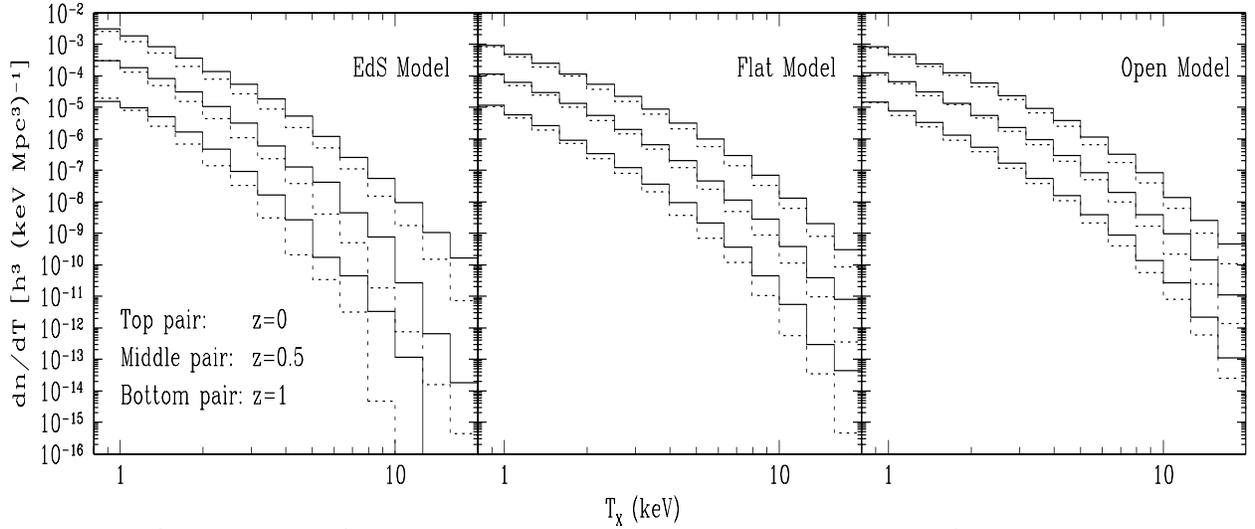}
\caption{Comparison of the boosted and non-boosted temperature functions given
by the merger trees for the three cosmologies we consider.
The histograms show the boosted (solid lines) and unboosted (dotted
lines) TFs for $z=0$ (top pair of lines),
$z=0.5$ (middle pair of lines), and $z=1$ (bottom pair of lines).
The lines that correspond to $z=0.5$ have been shifted down by one decade
and those at $z=1$ by two decades so that the different epochs are clearly
separated on the graph.
{\it Left:}\ EdS model.
{\it Middle:}\ Flat model.
{\it Right:}\ Open model.
\label{fig:dndT_tree}}
\end{figure}

\begin{figure}
\vskip3truein
\includegraphics{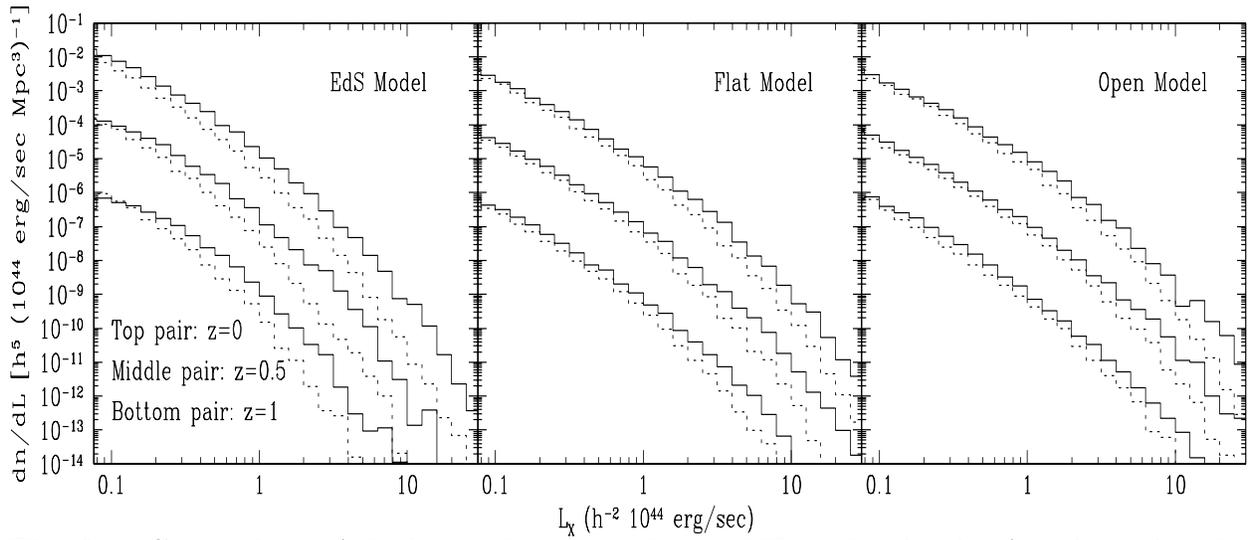}
\caption{Comparison of the boosted and non-boosted X-ray luminosity
functions given by the merger trees.
The notation is the same as in Figure~\protect\ref{fig:dndT_tree},
except that all $z=0.5$ data has been shifted down by two decades and
all $z=1$ data by four decades.
\label{fig:dndL_tree}}
\end{figure}

\begin{figure}
\vskip3truein
\includegraphics{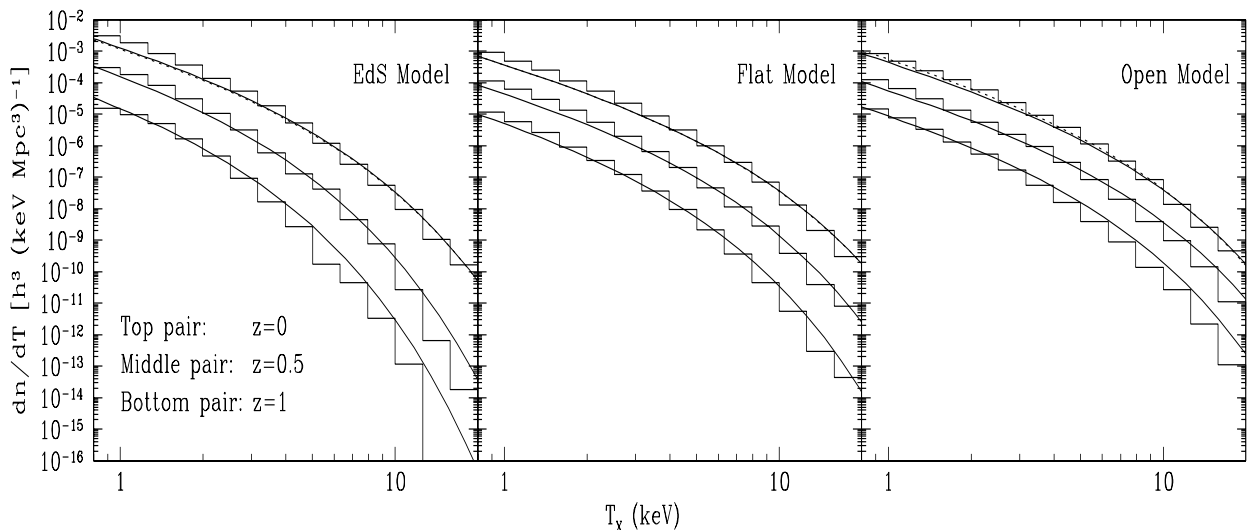}
\caption{Comparison of the boosted temperature functions given
by the merger trees (histograms) with the best-fit PS model
(smooth curves, equation~\protect\ref{eq:dndT}).  The smooth curve
shown for $z=0.5$ uses the values of $\Omega_0$ and $\sigma_8$
obtained from fitting the $z=0,0.5$ merger tree temperature functions
simultaneously, while the $z=1$ curve assumes the values obtained from
fitting $z=0,1$ simultaneously (see Table~\ref{tab:temp_fit}).  At
$z=0$ we show two smooth curves, one for each set of parameters.  The
solid line uses the parameters obtained from the $z=0,0.5$ fit and
the dashed line uses the parameters from the $z=0,1$ fit.  In most
cases the solid and dashed lines are indistinguishable.  As in
Figure~\ref{fig:dndT_tree}, all $z=0.5$ data have been shifted down by
one decade and all $z=1$ data by two decades.
\label{fig:dndT_fit}}
\end{figure}

\begin{figure}
\vskip3truein
\includegraphics{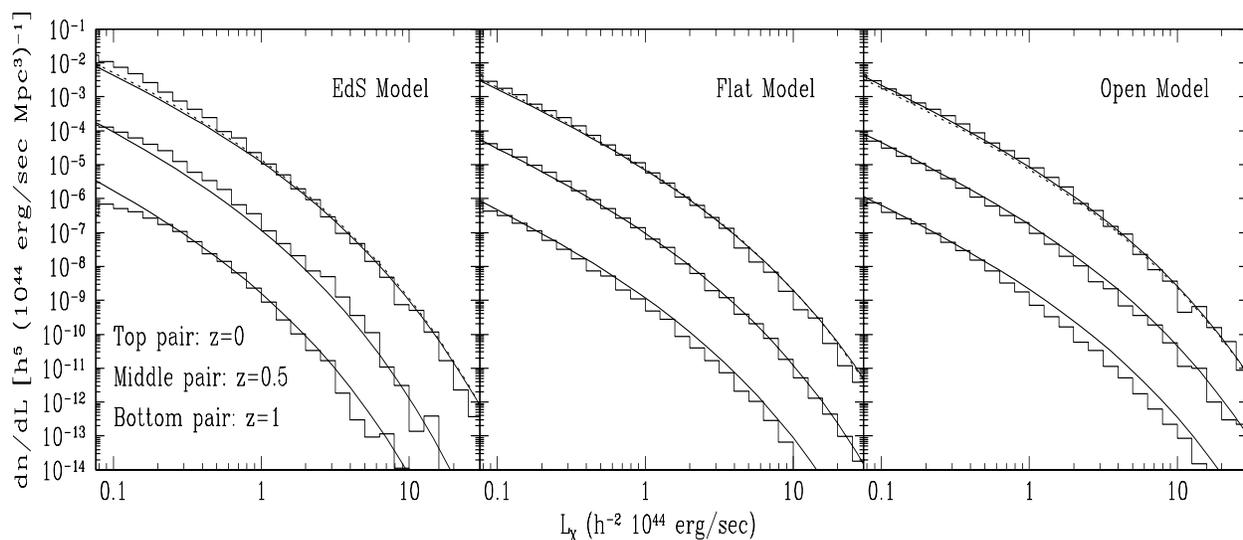}
\caption{Comparison of the boosted X-ray luminosity functions given
by the merger trees (histograms) with the best-fit PS model
(smooth curves).
The notation is the the same as in Figure~\protect\ref{fig:dndT_fit},
except that all $z=0.5$ data has been shifted down by two decades and
all $z=1$ data by four decades.
\label{fig:dLdT_fit}}
\end{figure}

As described in \S~\ref{sec:hydro_fit}, we wish to fits the portions
of the histograms shown in Figure~\ref{fig:histos} which correspond to
strong boosts.
We find that these portions of the cumulative time distribution are
well described by hyperbolas of the form
\begin{equation}   \label{eq:hyperbola}
\ln \left( \frac{t}{t_{\rm sc}} \right) =
\sqrt{
\left\{
\left[
\frac{T_X}{T_X (0)} - \left( \frac{T_X}{T_X (0)} \right)_{\rm max}
\, \right]^2 - 1 \right\}
\left( \epsilon_T^2 - 1 \right) }
 - \ln \left( \frac{t}{t_{\rm sc}} \right)_T \, .
\end{equation}
The parameters in this function are the largest temperature boost during the
merger $[ T_X / T_X (0) ]_{\rm max}$,
the duration of the period of maximum temperature boost
$(t / t_{\rm sc} )_T$,
and
$\epsilon_T$ which describes the eccentricity of the hyperbola.
As similar expression fits the time distribution of luminosity boosts,
with parameters
$[ L_X / L_X (0) ]_{\rm max}$,
$(t / t_{\rm sc} )_L$,
and
$\epsilon_L$.

After fitting the function given by equation~(\ref{eq:hyperbola}) to
each of the luminosity and temperature boost histograms shown in
Figure~\ref{fig:histos}, we determined the dependence of the boost
histogram parameters on the kinematics of the merger.
We characterize the merger kinematics by two dimensionless parameters.
The first is the fractional mass increase in the merger, $f_M$.
Following the discussion in above, let $M_<$ be the smaller of the two
masses of the merging subclusters and $M_>$ be the larger of the
two masses.
Then, $f_M \equiv  M_</(M_< + M_>)$.
Our second parameter $b'$ is the impact parameter $b$ for the merger,
divided by the sum of the core radii of the two merging subclusters,
$b' = b/(r_{c>} + r_{c<})$.
Here, $r_{c>}$ and $r_{c<}$ are the core radii of the merging subclusters,
which are taken to be 1/2 of the scaling radius $r_{s}$ for each subcluster.

We find that the variation of the boost histogram parameters with
the merger kinematics can be fit with the following functions (for the
temperature boost):
\begin{equation} \label{eq:tmaxfit}
\left[ \frac{T_X}{T_X (0)} \right]_{\rm max} ( f_M , b' ) =
\frac{A f_M^B }{C + b^{\prime 2}} + 1 \, ,
\end{equation}
\begin{equation} \label{eq:efit}
\epsilon ( f_M, b^{\prime} ) =
\left(
\frac{D f_M^E}{F + b^{\prime 2}} \right)^{-1}
\, ,
\end{equation}
\begin{equation} \label{timefit}
\ln \left( \frac{t}{t_{\rm sc}} \right) =
G \frac{ \ln(M_< + M_>) - H \ln(M_<^{1/3} + M_>^{1/3})}
{I + b^{\prime 2}} \, ,
\end{equation}
where $M_<$ and $M_>$ are in units of $M_\odot$.
Similar expressions are used for the X-ray luminosity boost.
The best-fit parameters in these fits are listed in Table~\ref{tab:fits}.

Note that equation~(\ref{eq:tmaxfit}) implies that the temperature and
luminosity boosts are largest for equal mass mergers ($f_M = 0.5$), and go
to zero as the mass of the smaller subcluster goes to zero
($f_M \rightarrow 0$).
Similarly, equation~(\ref{eq:efit}) implies that the histogram becomes
a vertical line as $f_M \rightarrow 0$;
that is, all of the time is spent with no boost.
As the mass fraction increases, $\epsilon$ decreases, and the boost
histogram becomes more extended toward large boosts.
The dependence of the maximum boost (equation~\ref{eq:tmaxfit}) and
eccentricity (equation~\ref{eq:efit}) on impact parameter is based on
the assumption that the gas distributions in clusters are well-fit by
the standard beta model
(Cavaliere \& Fusco-Femiano 1976),
and that the strength of the boost depends on the degree of interactions
of the densest portions of the two subclusters.
Thus, equations~(\ref{eq:tmaxfit}) \& (\ref{eq:efit}) give the largest
boost for a nearly head-on collision, and the boost goes to zero as the
impact parameter increases.
The quadratic dependence on $b^{\prime}$ reflects the form of the density
dependence in the beta model.

The form we adapted for the variation in the time duration of the merger
boost (equation~\ref{timefit}) follows from simple scaling for the
time scales for mergers.
Assuming that the two subclusters have fallen from a large distance to
a separation $d$, their relative velocity is
\begin{equation} \label{eq:vrel}
v_{\rm rel}^2 \approx
\frac{2G(M_< + M_>)}{d} \, ,
\end{equation}
where we approximate the subclusters as point masses.
The time scale for the merger boost is given by
$t_{\rm boost} \approx d / v_{\rm rel}$.
For a given formation epoch, the core radii and other length scales
associated with clusters scale as $r_c \sim M^{1/3}$.
Thus, we assume that the subcluster separation during the time of
maximum merger boost scales as
$d \sim M_<^{1/3}+M_>^{1/3}$.
One might expect that boost time to scale roughly as
\begin{equation} \label{eq:tboost}
t_{\rm boost} \sim \frac{d^{3/2}}{(M_< + M_>)^{1/2}}
\sim \frac{(M_<^{1/3}+M_>^{1/3})^{3/2}}{(M_<+M_>)^{1/2}} \, .
\end{equation}
The virial relation for the temperature of the gas in clusters leads to
$T \sim M^{2/3}$, which implies that the sound-crossing time
$t_{\rm sc}$ is nearly independent of mass.
These arguments determine the mass dependence of equation~(\ref{timefit}).
If these arguments held exactly, we would expect $H = 3$.
Although the best-fit values are close to this (Table~\ref{tab:fits}),
the fits are improved if $H$ is allowed to vary.
We adapt the same dependence on $b^{\prime}$ as in the other fits.

Ultimately these fits will be used to determine the luminosity and
temperature boosts associated with a pair of merging halos given the
time during the merger at which the merging cluster is observed.
For a given value of $( t / t_{\rm sc} )$,
equations~(\ref{eq:hyperbola}), (\ref{eq:tmaxfit}), (\ref{eq:efit}), \&
(\ref{timefit}) and the parameters in Table~\ref{tab:fits} reproduce
the temperature boosts $T_X / T_X (0)$ with an average accuracy of better
than 2\% for all boosts greater than 1.5.
The accuracy of the fit typically increases with
the strength of the boost; thus the accuracy for the strongest
temperature boosts are generally better than 1\%.
The fits for the luminosity boosts $L_X / L_X (0)$ have an average
accuracy better than 7\% for boosts greater than 1.5.

\clearpage

\begin{deluxetable}{lccccc}
\tablewidth{6truein}
\tablecaption{Increase in Number of Hot, Luminous Clusters Due
to Merger Boosts
\label{tab:boosts}}
\tablehead{
\colhead{}&
\colhead{}&
\multicolumn{4}{c}{Fractional Increase in Number of Clusters}\\
\colhead{Model}&
\colhead{$z$}&
\colhead{$(T > 7$ keV)}&
\colhead{$(T > 10$ keV)}&
\colhead{$(L_X > 7 \times 10^{43})$}&
\colhead{$( L_X > 5 \times 10^{44})$}}
\startdata
Open&0.0&\phn1.96&\phn2.35&1.97&\phn3.43\\
Open&0.5&\phn2.24&\phn4.64&1.71&\phn4.97\\
Open&1.0&\phn2.32&\phn3.30&1.71&\phn3.91\\
Flat&0.0&\phn2.10&\phn2.24&2.01&\phn4.37\\
Flat&0.5&\phn2.56&\phn3.54&2.04&\phn8.11\\
Flat&1.0&\phn3.37&\phn9.49&2.05&\phn8.90\\
EdS &0.0&\phn2.73&\phn5.62&4.08&20.7\phn\\
EdS &0.5&11.3\phn&37.7\phn&4.51&24.1\phn\\
EdS\tablenotemark{a} &1.0&\phn3.25&\phn6.72&2.65&21.7\phn\\
\enddata
\tablenotetext{a}{For the EdS model at $z=1$, we use minimum
values of $T > 2$ keV, $T > 5$ keV, $L_X > 5 \times 10^{43} \,
h^{-2}$ erg s$^{-1}$, and $L_X > 2 \times 10^{44} \, h^{-2}$ erg s$^{-1}$.}
\end{deluxetable}

\begin{deluxetable}{lcrrrrrrrrr}
\tablewidth{5truein}
\tablecaption{Fitted Parameters for Merger Tree Temperature Function
\label{tab:temp_fit}}
\tablehead{
\colhead{Model}&
\colhead{Redshifts}&
\colhead{$\sigma_{8, \rm a}$}&
\colhead{$\Omega_{0, \rm a}$}&
\colhead{$\sigma_{8, \rm nb}$}&
\colhead{$\Omega_{0, \rm nb}$}&
\colhead{$\sigma_{8, \rm b}$}&
\colhead{$\Omega_{0, \rm b}$}}
\startdata
Open&0,0.5&0.8274&0.3&0.913&0.26&1.190&0.13\\
Open&0,1.0&0.8274&0.3&0.912&0.26&1.046&0.21\\
Flat&0,0.5&0.8339&0.3&0.919&0.25&1.096&0.20\\
Flat&0,1.0&0.8339&0.3&0.905&0.26&1.095&0.20\\
EdS &0,0.5&0.5138&1.0&0.555&0.86&0.667&0.71\\
EdS &0,1.0&0.5138&1.0&0.575&0.77&0.693&0.63\\
\enddata
\tablecomments{The subscript ``a'' indicates the actual values used to
construct the trees, ``nb'' indicates the fitted parameters for the
unboosted temperature merger tree functions, and ``b'' indicates the fitted
parameters to the boosted temperature functions.
}
\end{deluxetable}

\begin{deluxetable}{lcrrrrrrrrr}
\tablewidth{5truein}
\tablecaption{Fitted Parameters for Merger Tree Luminosity Function
\label{tab:lum_fit}}
\tablehead{
\colhead{Model}&
\colhead{Redshifts}&
\colhead{$\sigma_{8, \rm a}$}&
\colhead{$\Omega_{0, \rm a}$}&
\colhead{$\sigma_{8, \rm nb}$}&
\colhead{$\Omega_{0, \rm nb}$}&
\colhead{$\sigma_{8, \rm b}$}&
\colhead{$\Omega_{0, \rm b}$}}
\startdata
Open&0,0.5&0.8274&0.3&1.057&0.49&1.113&0.38\\
Open&0,1.0&0.8274&0.3&1.085&0.59&1.145&0.46\\
Flat&0,0.5&0.8339&0.3&0.857&0.19&1.037&0.37\\
Flat&0,1.0&0.8339&0.3&0.849&0.17&1.000&0.23\\
EdS &0,0.5&0.5138&1.0&0.535&0.82&0.655&1.20\\
EdS &0,1.0&0.5138&1.0&0.546&0.99&0.638&0.93\\
\enddata
\end{deluxetable}

%
%
\begin{deluxetable}{lrrrrrrrrr}
\tablewidth{5truein}
\tablecaption{Fitting Parameters for Merger Boost Histograms
\label{tab:fits}}
\tablehead{
\colhead{Boost}&
\colhead{A}&
\colhead{B}&
\colhead{C}&
\colhead{D}&
\colhead{E}&
\colhead{F}&
\colhead{G}&
\colhead{H}&
\colhead{I}}
\startdata
$T_X / T_X (0)$&195&0.448&49&132&0.539&127&349    &2.81& 94\\
$L_X / L_X (0)$&240&0.659&29& 92&0.316& 84&$-$96&3.29&129\\
\enddata
\end{deluxetable}

\clearpage

\clearpage

\clearpage

\end{document}